# Frequency stabilization and noise-induced spectral narrowing in resonators with zero dispersion

L. Huang[1,2], S. M. Soskin[3,4], I. A. Khovanov[5], R. Mannella[6], K. Ninios[7], and H. B. Chan[1,2]*

[1]Department of Physics, the Hong Kong University of Science and Technology, Clear Water Bay, Kowloon, Hong Kong, China.

[2] William Mong Institute of Nano Science and Technology, the Hong Kong University of Science and Technology, Clear Water Bay, Kowloon, Hong Kong, China.

[3]Institute of Semiconductor Physics, National Academy of Sciences of Ukraine, 03028 Kiev, Ukraine

[4]Department of Physics, Lancaster University, Lancaster LA1 4YB, United Kingdom

[5]School of Engineering, University of Warwick, Coventry CV4 7AL, United Kingdom

[6]Dipartimento di Fisica, Universita di Pisa, 56127 Pisa, Italy

[7]Department of Physics, University of Florida, Gainesville, Florida 32611, USA

*email: hochan@ust.hk

Abstract

**Mechanical resonators are widely used as precision clocks and sensitive detectors that rely on the stability of their eigenfrequencies. The phase noise is determined by different factors including thermal noise, frequency noise of the resonator and noise in the feedback circuitry. Increasing the vibration amplitude can mitigate some of these effects but the improvements are limited by nonlinearities that are particularly strong for miniaturized micro- and nano-mechanical systems. Here we design a micromechanical resonator with non-monotonic dependence of the eigenfrequency on energy. Near the extremum, where the dispersion of the eigenfrequency is zero, the system regains certain characteristics of a linear**

**resonator, albeit at large amplitudes. The spectral peak undergoes narrowing when the noise intensity is increased. With the resonator serving as the frequency-selecting element in a feedback loop, the phase noise at the extremum amplitude is ~ 3 times smaller than the minimal noise in the conventional nonlinear regime.**

**Introduction**

Resonant detectors and frequency standards typically operate in the linear regime, where the eigenfrequency $\omega$ is independent of the amplitude and hence the energy. The frequency stability depends on the width of the spectral peak, which is in turn proportional to the damping constant. For systems coupled to a thermal bath, the damping constant is related to thermal fluctuations via the fluctuation-dissipation theorem. The spectral width is an important parameter in mesoscopic resonators including nanomechanical devices[1-6], Josephson junctions[7, 8] and optomechanical systems[9]. Apart from practical implications on the phase noise, the spectral width also provides useful information on the underlying relaxation mechanisms[10-12]. Recently, there has been much interest in different factors that modify the spectral response of mesoscopic resonators. It has been observed in nanomechanical resonators that spectral broadening can arise from the fluctuations of the eigenfrequency[13-16]. Alternatively, the interplay between nonlinearity, fluctuations and relaxation has also been found to have a profound effect on the spectral lineshape[17, 18].

Mechanical oscillators used in sensing or time-keeping often rely on a feedback circuit to maintain self-sustained oscillations[19-22] so that the eigenfrequency of the resonator can be tracked. A number of factors contribute to the phase noise, including the spectral width of the resonator and noise generated by the feedback circuit. These effects can be mitigated by increasing the vibration amplitude, provided that resonator remains in the linear regime. The improvement in frequency stability, however, is limited by nonlinear effects. When the vibration amplitude increases beyond the linear regime, the eigenfrequency is no longer independent of energy. The nonlinearity converts amplitude fluctuations into frequency fluctuations, leading to degradations in phase stability[21]. Such nonlinear effects are particularly strong in miniaturized silicon micro- and nano-mechanical devices that have become viable candidates to replace bulky quartz resonators commonly used as frequency references[23]. Stabilization of the frequency in the presence of strong nonlinearity has attracted much interest recently. For example, one approach makes use of higher order modes of the resonator. At amplitudes where the two modes are in internal resonance[21, 24-26], the frequency fluctuations of the lower mode are shown to be significantly reduced[21].

On a more fundamental level, the nonlinearity modifies the spectral lineshape of resonators. When resonators are subjected to thermal and/or external fluctuations, the noise creates a distribution of energies that, together with the nonlinearity, tends to broaden the peak in the power spectrum[18, 27]. The overall shape of the spectral peak is determined by the interplay of the frequency straggling due to nonlinearity and the frequency uncertainty associated with decay. As the noise intensity increases, the peak broadens, generally, and becomes asymmetric. Changes in spectral peaks with increasing temperature were observed in the optical spectra of localized and resonant vibrations in solids[28]. However, a full quantitative comparison with theory was difficult due to uncertainties in system parameters and difficulties in controlling them.

In this work we design an electromechanical resonator with non-monotonic dependence of the eigenfrequency $\omega$ on energy $E$ to yield improved spectral width and frequency stability. At the extremum energy $E_{zd}$, the dispersion of $\omega(E)$ is zero so that $d\omega/dE = 0$. Our system opens the possibility for the investigation of a wealth of phenomena in "zero-dispersion" systems[29-40] that are associated with the large contribution of vibrations over a range of energies in the vicinity of $E_{zd}$ occurring at almost the same frequency. These phenomena bear resemblance to van Hove singularities in the density of states of solids[41, 42]. In our mechanical resonator, the non-monotonic dependence of $\omega$ on energy $E$ is achieved by tuning the negative nonlinearity induced by electrostatic force from a nearby electrode relative to the positive intrinsic nonlinearity of the springs. We demonstrate that in the zero-dispersion regime, the peak of the fluctuation spectrum undergoes substantial narrowing as the noise intensity is increased, in good agreement with theory. We also set our devices into self-sustained oscillations with active feedback. The standard deviation of the oscillation frequency at the optimal vibration amplitude is demonstrated to be a factor of ~3 smaller in the zero-dispersion regime compared to the minimal noise in the conventional nonlinear regime. To our knowledge, zero-dispersion phenomena have not been directly observed in physical systems other than analog circuits and computer simulations[30, 32]. Our findings demonstrate that zero-dispersion phenomena can play an important role in micro/nano mechanical systems, potentially leading to new methods of detection and frequency stabilization.

# Results

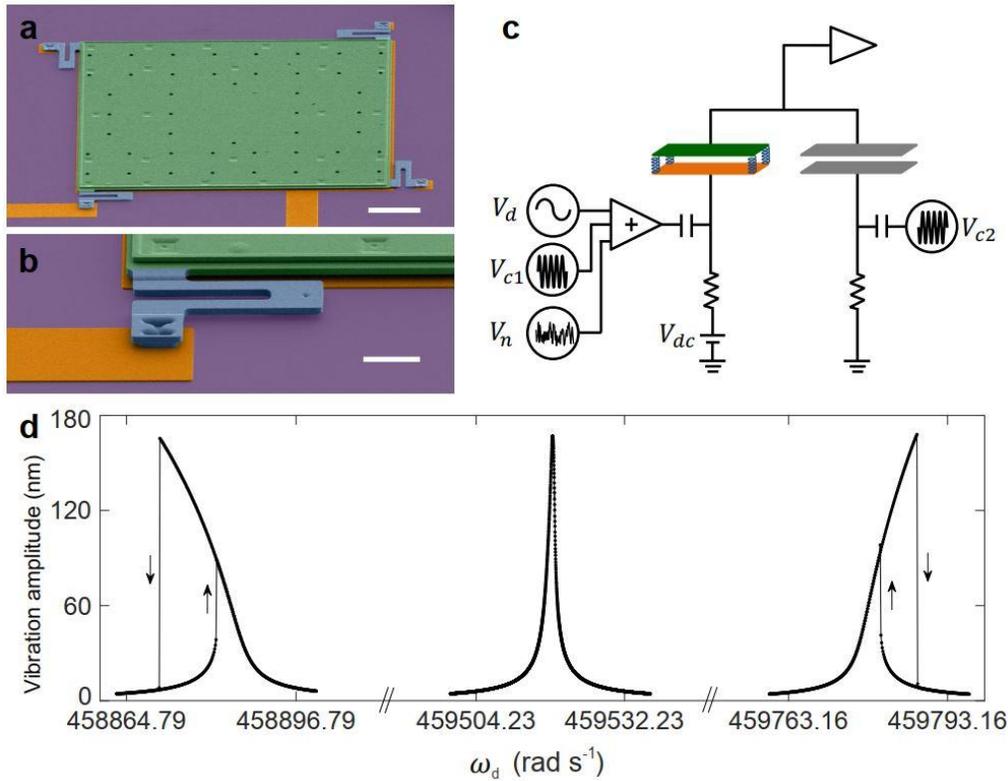

**Fig. 1** Micromechanical trampoline resonator. **a** Colorized scanning electron micrograph of the resonator that consists of a movable silicon top plate (green) supported by serpentine springs (blue) at its four corners. The white scale bar measures 50 μm. **b** A close-up on one of the four springs at a corner. The scale bar measures 15 μm. **c** Simplified schematic showing the electrostatic actuation and detection of vibrations of the top plate using a capacitance bridge. **d** The electrostatic contribution to the Duffing nonlinearity can be controlled by $V_{dc}$. Measured vibration amplitude is plotted as a function of driving frequency for a fixed driving force amplitude and three different $V_{dc}$: -2.3V, -1.6V and -1.2V from left to right, where the Duffing nonlinearity changes from positive to near zero to negative.

**Micromechanical trampoline resonator.** Figure 1a shows the micromechanical trampoline resonator used in our experiment. The device consists of a highly-doped polycrystalline silicon plate (3.5 μm × 300 μm × 300 μm) suspended by springs at its four corners (Fig. 1b). Electrical voltage $V$ is applied to a fixed electrode of the same size underneath the top plate to excite vibrations of the top plate in the fundamental mode that involves translational motion normal to the substrate. Measurements are performed at 4 K and $10^{-6}$ Torr. The top plate can be modeled as a resonator in one dimension:

$$\ddot{q} = -2\Gamma \dot{q} + \frac{1}{m}[F_s(q) + F_e(q)] \tag{1}$$

where $q$ is the plate displacement, $m$ is the mass, $\Gamma$ (0.496 rad s$^{-1}$) is the damping constant. $F_s$ is the restoring force of the spring:

$$F_s(q) \approx -m\omega_s^2 q - m\beta_s q^3 - m\mu_s q^5. \tag{2}$$

The spring nonlinearity coefficients $\beta_s$ and $\mu_s$ are positive because of the increase in tension with displacement $q$. Asymmetry of the spring is negligible at relevant scales of $q$. $F_e(q)$ is the electrostatic force:

$$F_e(q) \equiv F_e(q, V^2) = \frac{1}{2}\frac{dC(q)}{dq}V^2, \tag{3}$$

where $C = \varepsilon S/(g-q)$ is the capacitance between the plate and the electrode, $\varepsilon$ is the permittivity of free space, $S$ is the area of the plate, and $g$ is the initial gap of 2 μm between the plate and the electrode.

The applied voltage $V$ between the plate and the electrode includes up to four components:

$$V = V_{dc} + V_d(t) + V_n(t) + V_{c1}(t) \tag{4}$$

$V_{dc}$ is the dc voltage. $V_d(t) = V_{ac}\cos(\omega_d t)$ (with $V_{ac} \ll |V_{dc}|$) is the ac driving at frequency $\omega_d$ close to the eigenfrequency of the fundamental mode. It leads to a resonant periodic electrostatic force on the plate. $V_n(t)$ is the noise voltage that generates a random force. The last term $V_{c1}(t) = V_c\cos(\omega_c t)$ represents a sinusoidal carrier voltage at amplitude $V_c = 300$ mV and frequency $\omega_c = 2.5 \times 10^7$ rad s$^{-1}$. It is used to measure the change in capacitance between the plate and the electrode, from which $q$ can be inferred (see Methods). For all results shown in this paper, $V_{ac} \ll V_c \ll |V_{dc}|$. Besides, since $\omega_c \gg \omega_d$, the only effect of the carrier voltage on the dynamics of $q$ is a slight increase of the effective value of the dc voltage $\tilde{V}_{dc}^2 = V_{dc}^2 + V_c^2/2$.

**Dependence of frequency of eigenoscillations on amplitude.** One of the most important characteristics of the resonator is the dependence of the frequency of eigenoscillation of the resonator $\omega$ on its amplitude $a$: it determines characteristic features of dynamics both in case of resonant periodic driving[43, 44] and in case of driving by noise[27, 38]. The dependence of the electrostatic force on $V_{dc}$ allows us to control $\omega(a)$ and turn the zero dispersion behavior on or off. Such control is particularly efficient in our resonator because the gap $g$ is much smaller than the characteristic distance scale of nonlinearity in the spring restoring force $L_n = \omega_s/\sqrt{\beta_s}$. Apart from the efficient control, it also allows us to develop for $\omega(a)$ an asymptotic theory with a small parameter $\tilde{v}$ that is proportional to $\tilde{V}_{dc}^2$:

$$\tilde{v} \equiv \tilde{v}\left(\tilde{V}_{dc}^2\right) = \frac{\varepsilon S}{2mg^3\omega_s^2}\tilde{V}_{dc}^2. \tag{5}$$

As it will be shown below, the range of $\tilde{V}_{dc}^2$ relevant to our experiment is

$$\tilde{V}_{dc}^2 \sim \beta_s mg^5/(2\varepsilon S). \tag{6}$$

Therefore, $\tilde{v}$ in this range is of the order of $[g/(2L_n)]^2$, which is $\ll 1$. As we will later explain, $\tilde{v}$ plays a crucial role in determining whether the resonator exhibits zero-dispersion.

At the equilibrium position $q_{eq} \approx \tilde{v}g \ll g$, the spring and electrostatic forces balance each other. We consider the force $F = F_s + F_e$ as a function of $x \equiv q - q_{eq}$ and expand $F(x)$ into Taylor series up to the 5$^{th}$ order (see Supplementary Note 1):

$$F(x) \equiv F(x, \tilde{V}_{dc}^2) = m \sum_{n=1}^{5} \alpha_n x^n, \tag{7}$$

where the odd coefficients contain contributions from both the spring and the electrostatic forces:

$$\alpha_1 \approx \omega_s^2 (1 - 2\tilde{v}), \tag{8}$$

$$\alpha_3 \approx \beta_s - \frac{4\omega_s^2}{g^2}\tilde{v} \equiv \beta_s \left\{1 - \frac{\tilde{v}}{[g/(2L_n)]^2}\right\}, \tag{9}$$

$$\alpha_5 \approx \mu_s - \frac{6\omega_s^2}{g^4}\tilde{v} \equiv -\frac{6\omega_s^2}{g^4}\tilde{v}\left\{1 - \frac{[g/(2L_n)]^2}{\tilde{v}} \frac{8\mu_s L_n^2}{3\beta_s}\left(\frac{g}{2L_n}\right)^2\right\} \tag{10}$$

(relatively small terms proportional to $\tilde{v}^n$ with $n \geq 2$ are neglected in Eqs. (8)-(10)). In contrast, the even coefficients contain no contributions from the spring [see Eq. (2)]. They originate exclusively from the electrostatic force and, to the lowest order in $\tilde{v}$, they are proportional to $\tilde{v}\omega_s^2$, namely $\alpha_2 \approx 3\tilde{v}\omega_s^2/g$ and $\alpha_4 \approx 5\tilde{v}\omega_s^2/g^3$.

Most nonlinear micro- and nano-mechanical resonant systems can be modeled as Duffing oscillators that includes up to the cubic nonlinear term. Obtaining zero dispersion phenomena in our system requires keeping nonlinear terms up to at least the fifth order. They generate a fourth order term in the dependence of $\omega$ on the vibration amplitude $a$, in addition to the usual quadratic term as in a Duffing oscillator (Supplementary Note 1):

$$\omega \approx \omega_1 + \kappa a^2 + \eta a^4, \tag{11}$$

where

$$\omega_1^2 \approx \omega_s^2 (1 - 2\tilde{v}), \tag{12}$$

$$\kappa \approx \frac{3\beta_s}{8\omega_s}\left\{1 - \frac{\tilde{v}}{[g/(2L_n)]^2}\right\}, \tag{13}$$

$$\eta \approx -\frac{15\omega_s \tilde{\nu}}{8g^4}. \tag{14}$$

In contrast to $\eta$ (Eq. (14)) which is necessarily negative, $\kappa$ (Eq. (13)) can be chosen to be either positive or negative by controlling $\tilde{\nu}$ via $\tilde{V}_{dc}^2$ (Supplementary Note 1). When the signs of $\kappa$ and $\eta$ are opposite, the eigenfrequency possesses a parabolic maximum as a function of $a^2$, providing a versatile platform for investigating zero dispersion phenomena.

**Resonance response curves.** We study the dependence of the vibration amplitude on the frequency of a sinusoidal drive when $V_n(t)$ is set to zero. The equation of motion can thus be simplified to:

$$\ddot{x} + 2\Gamma\dot{x} + \omega_1^2 x + \alpha_3 x^3 + \alpha_5 x^5 = (F_{ac}/m)\cos(\omega_d t) \tag{15}$$

where $F_{ac} = \frac{\epsilon S}{g^2}|V_{dc}|V_{ac}$. Contributions from $\alpha_i$ with even $i$ are negligible (See Supplementary Note 1). The results for small $F_{ac}$ are plotted as resonance response curves in Fig. 1d. For small $F_{ac}$, the vibration amplitude $A$ is also small such that the term $\alpha_5 x^5$ in Eq. (15) and the term $\eta a^4$ in Eq. (11) for $a = A$ can be neglected. The system reduces to the periodically driven underdamped Duffing oscillator then. Figure 1d shows the electrostatic spring softening effect: the resonance peaks shift to lower frequency as $V_{dc}^2$ increases, as described by Eq. (12). Apart from the frequency shift, the nonlinear coefficients also change with $V_{dc}^2$. At small $V_{dc}^2$, the inherently positive spring contribution to the cubic nonlinearity $\alpha_3$ strongly dominates over the negative contribution from the electrostatic force, so that the resonance curve bends towards high frequencies and exhibits clockwise hysteresis[45, 46]. While the spring contribution is fixed, the absolute value of the negative electrostatic contribution increases in proportion to $\tilde{V}_{dc}^2$.

At $V_{dc}$ = -1.6 V, it balances the spring contribution so that $\alpha_3 \sim 0$. As a result, the asymmetry in the resonance peak is largely gone and the peak is well-fitted by the response of a damped harmonic oscillator. As $V_{dc}^2$ further increases, $\alpha_3$ becomes negative so that the resonance peak bends towards low frequencies. A somewhat similar evolution of the resonance response curves was observed in Refs. [47, 48] but the reverse of the curve bending was caused there primarily by the strong growth of the $\alpha_2$ contribution, in contrast to our case. Regardless of the origin of such an evolution of resonance curves, the regime in which $\kappa$ becomes zero may be used to increase the dynamic range of micromechanical devices[48].

Figure 2a shows measured and calculated (see Supplementary Note 1 for details) resonance curves at $V_{dc}$ = -2.3 V, where the nonlinearity due to the electrostatic force dominates over the spring nonlinearity, for driving amplitudes of 36.7 pN and 202 pN. As the driving amplitude increases, the hysteresis loop widens and the frequency $\omega_{peak}$ with the largest vibration amplitude $A_{peak}$ shifts further towards low frequencies. By measuring the resonance curves at many values of the driving amplitude and identifying the peak vibration amplitude, we obtain the backbone line $A_{peak}(\omega_{peak})$. The reversed function $\omega_{peak}(A_{peak})$ provides a good approximation of the frequency of eigenoscillations (see Supplementary Note 1) with amplitude $a$ equal to $A_{peak}$[43]. Figure 2b plots the dependence of $\omega_{peak}$ on $A_{peak}^2$, which we call the "transformed backbone line". The solid line represents a fit using Eq. (11) with $a = A_{peak}$, yielding $\omega_l = \omega_{la}$ = 458874.45 rad s$^{-1}$, $\kappa$ = - 2.16 × 10$^{14}$ rad s$^{-1}$ m$^{-2}$ and $\eta$ = - 1.13 × 10$^{26}$ rad s$^{-1}$ m$^{-4}$.

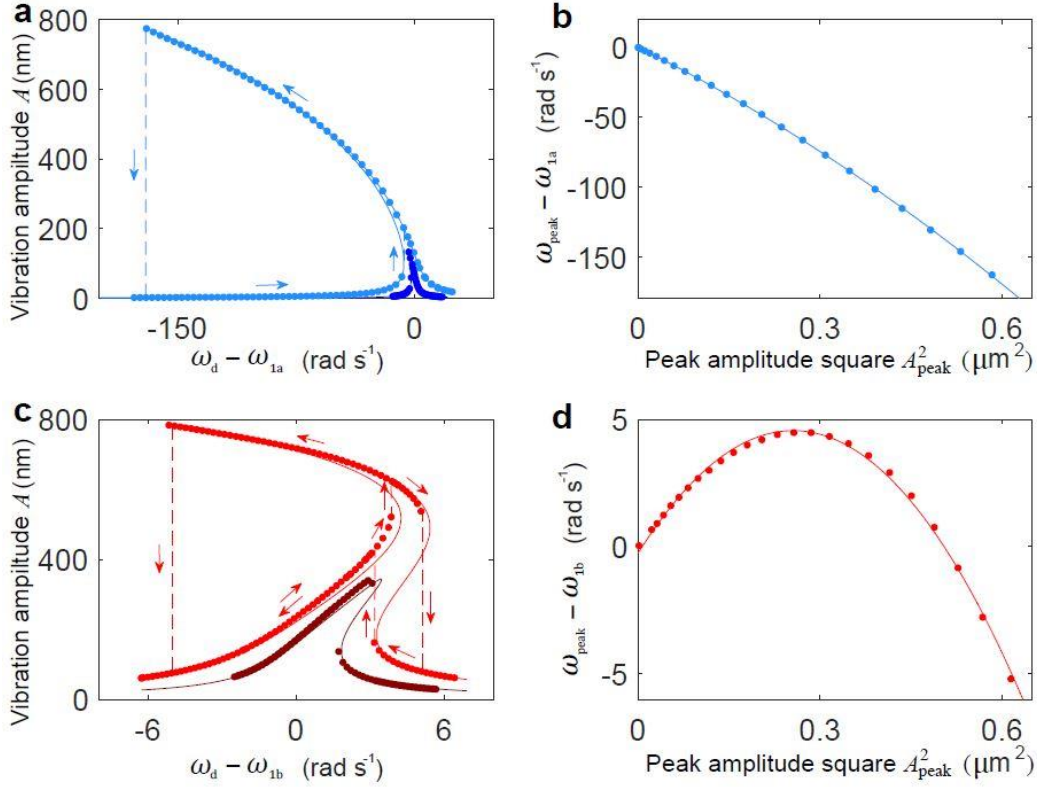

**Fig. 2** Resonance curves and transformed backbone lines in the conventional nonlinear and zero dispersion regimes. **a** At $V_{dc}$ = -2.3 V, the resonator is in the conventional nonlinear regime. The resonance curve bends towards lower frequencies as the driving amplitude $F_{ac}$ is increased from 36.7 pN (dark blue) to 202 pN (blue). Measurement and theory (Supplementary Note 1) are represented by circles and lines respectively. **b** Measured transformed backbone line (circles) showing the shift in $\omega_{peak}$ relative to the natural frequency vs. the squared vibration amplitude at the peak of the resonance curve as the driving amplitude increases. The solid line is a quadratic fit. **c** At $V_{dc}$ = -1.51 V, the resonator is in the zero dispersion regime. The resonance curve for $F_{ac}$ = 75.2 pN (dark red) bends to higher frequencies. For $F_{ac}$ = 171 pN (red), the initial bending is also towards higher frequencies but as the vibration amplitude further increases, the bending changes direction. **d** Accordingly, the transformed backbone line becomes non-monotonic and exhibits a maximum. The error bars are smaller than the dot size in all panels.

If $V_{dc}$ > -1.6 V, the transformed backbone line becomes non-monotonic, i.e. the device acquires zero-dispersion properties. Figure 2c shows resonance curves for two driving amplitudes at $V_{dc}$ = -1.51 V. At the lower driving amplitude, the curve bends towards higher frequencies due to the positive $\kappa$. The sign of $\kappa$ is changed compared to Fig. 2a because, with a smaller $V_{dc}^2$, the spring contribution dominates over the electrostatic force. In Eq. (13), the ratio $\tilde{v}/[g/(2L_n)]^2$ becomes slightly smaller than 1. At larger driving amplitude, the curve acquires a mixed behavior: it bends to higher frequencies at sufficiently low vibration amplitude $A$ but, as $A$ exceeds certain limit, the bending changes direction towards lower frequencies since the negative 4$^{th}$ power term in $\omega(A)$ [Eq. (11)] overcomes the positive quadratic term (Supplementary Note 1 explains why $\eta$ is negative in our experiment). There exists two separate frequency ranges with hysteretic behavior[49]. The system can have one, two or three co-existing stable vibration states depending on the excitation frequency, with a total of 4 bifurcation frequencies. Figure 2d shows (by circles) the measured non-monotonic transformed backbone line $\omega_{peak}(A_{peak}^2)$. The solid line represents a fit using Eq. (11) with $a = A_{peak}$, yielding $\omega_l = \omega_{lb} = 459572.9$ rad s$^{-1}$, $\kappa = 3.77 \times 10^{13}$ rad s$^{-1}$ m$^{-2}$ and $\eta = -7.37 \times 10^{25}$ rad s$^{-1}$ m$^{-4}$. Unlike Fig. 2b, $\kappa$ and $\eta$ have opposite signs. The transformed backbone line exhibits a local maximum, at which $d\omega/dE = 0$, that provides a platform for exploring and utilizing zero-dispersion phenomena.

**Noise-induced spectral widening and narrowing.** Next we investigate the spectral density of fluctuations of our resonator in the conventional Duffing and the zero dispersion regimes. The measurements are performed by injecting noise $V_n$ in the driving voltage [Eq. (4)] while turning off the ac drive $V_d$. Noise $V_n$ is Gaussian, centered around $\omega_l$ with a bandwidth of ~ 5000 rad s$^{-1}$ for the conventional nonlinear regime and ~2000 rad s$^{-1}$ for the zero-dispersion regime (see Methods). All the resonance and spectral peaks investigated (such as those in Fig. 1) lie well within this bandwidth, so that the noise can be assumed white in the context of the fluctuation spectrum, playing a role similar to thermo-mechanical noise (Supplementary Note 2). The dynamics is well-described by Eq. (15) with the following modifications: (i) the ac force $F_{ac}\cos(\omega_d t)$ being replaced by the auxiliary white noise $F_n^{(w)}$ with intensity $D_w$ defined in Supplementary Equation 3838 and (ii) $\tilde{V}_{dc}^2$ in the left-hand side being replaced by the slightly larger value as described in Supplementary Note 2.

The spectral density of fluctuations $\tilde{Q}(\omega)$ is defined as the half-Fourier transform of the correlation function of coordinate $Q(t)$[29-32, 36, 38, 50]:

$$Q(t) \equiv \langle (q(t) - \langle q \rangle)(q(0) - \langle q \rangle) \rangle, \tag{16}$$

$$\tilde{Q}(\omega) \equiv \frac{1}{\pi} \mathrm{Re}\left[\int_0^\infty dt Q(t)\exp(-i\omega t)\right]. \tag{17}$$

Details of the methods of measurements of $\tilde{Q}(\omega)$ and of its theoretical calculations are given in Supplementary Note 3. The procedure involves transformation from the description in terms of dynamical variables to that of the non-stationary conditional probability density and, accordingly, from the description by means of the Langevin equation to that by means of the Fokker-Planck equation (FPE)[50], with the further substantial simplification of the solution of the FPE and calculation of the spectrum using the method suggested in Ref. [30].

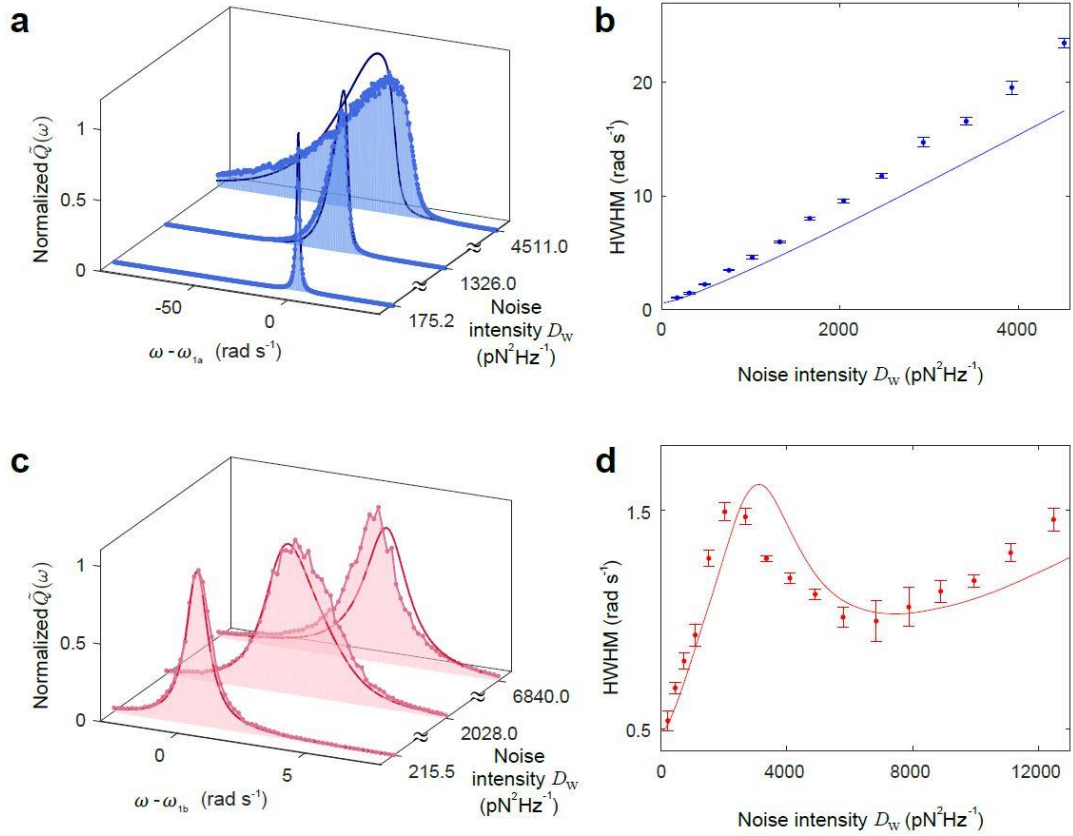

**Fig. 3** Spectral widening/narrowing in the conventional/zero-dispersion case. **a** In the conventional nonlinear regime ($V_{dc}$ = -2.3V), the spectral peak becomes wider and asymmetric as the noise intensity is increased. Each peak is normalized by its maximum measured value. **b** Measured (circles) and calculated (solid line) half-width at half-maximum (HWHM) of the spectral peaks increase monotonically with the noise intensity. **c** and **d** Similar plots for the zero dispersion regime for $V_{dc}$ = -1.51 V. Note the difference of scales as compared to **a** and **b** respectively. The HWHM shows a distinctly non-monotonic dependence on the noise intensity, yielding a range of noise intensity where the spectral peak becomes narrower as the noise intensity increases. Error bars represent ±1 s. e.

Figures 3a and 3c plot the measured spectral density of fluctuations (see Methods) at $V_{dc}$ of -2.3 V and -1.51 V, corresponding to the monotonic and non-monotonic transformed backbone lines in Fig. 2b and Fig. 2d respectively. When the noise intensity $D_w$ is small, the spectrum is Lorentzian as in a harmonic oscillator. The peak is symmetric, centered at the natural frequency and the width is solely determined by the damping constant $\Gamma$. As $D_w$ increases, the interplay between fluctuations and nonlinearity leads to profound changes in the width, height and shape of the peaks. The noise-induced energy straggling leads to a corresponding frequency straggling. Therefore, the width of the spectral peak no longer depends solely on the damping constant. As shown in Fig. 3a for the case where the transformed backbone line is monotonic, the width of the peak increases with $D_w$ and the peaks become asymmetric. Figure 3b shows that the measured change in the width of the peak is roughly proportional to the noise intensity. In Fig. 3a and Fig. 3b, the solid lines show results of the theoretical calculations with no fitting parameters. Given that the asymptotic theory is only of the lowest order, the agreement may be considered satisfactory.

When the system is in the zero dispersion regime at $V_{dc}$ = -1.51 V, there exists an energy $E_{zd}$ (with corresponding eigenfrequency $\omega_{zd} \equiv \omega(E_{zd})$) at which $d\omega/dE = 0$. In the limit when damping approaches zero, the spectral density of fluctuation is predicted to diverge at $\omega_{zd}$ [29] because vibrations in a relatively large range of energies take place with almost the same frequency. In our system, the ratio $(\omega_{zd} - \omega_l)/\Gamma$ is not sufficiently large to distinctly reveal this peak. Nevertheless, the reduced frequency straggling associated with $d\omega/dE = 0$ results in a decrease of the spectral width as the noise intensity is increased. Such noise-induced spectral narrowing is clearly seen in Fig. 3c, where the spectral peak at $D_w$ = 6840 pN$^2$ Hz$^{-1}$ has a smaller width compared to $D_w$ = 2028 pN$^2$ Hz$^{-1}$. Figure 3d plots the dependence of the spectral width on noise intensity. As $D_w$ increases, the width initially increases, attaining a maximum value at $D_w$ ~ 2030 pN$^2$ Hz$^{-1}$. Upon further increase in $D_w$, it drops by about 30% at $D_w$ ~ 6800 pN$^2$ Hz$^{-1}$. At even larger $D_w$, the spectral width resumes its increase. Similar to the conventional case, the theoretically calculated spectra demonstrate a reasonable agreement with the experiments with no fitting parameters.

**Frequency stabilization in the zero-dispersion regime**. The reduced spectral width in the zero dispersion regime offers new opportunities for frequency stabilization in applications of frequency standards. In the linear regime, the spectral width directly determines the frequency stability when the resonator is driven into self-sustained oscillations via active feedback, where the mechanical resonator serves as the frequency-selecting element[20, 21]. To evade thermal noise or noise from the feedback amplifier, it is beneficial to maximize the vibration amplitude of the mechanical resonator, provided that the response of the mechanical element remains linear. In practice, however, as the vibration amplitude increases, nonlinear effects become more prominent. Fluctuations in the vibration amplitude are converted via the nonlinearity into frequency fluctuations. Increasing the vibration amplitude eventually leads to a degradation of phase stability of self-sustained oscillations. To demonstrate this phenomenon, we drive our resonator into self-sustained vibrations using a phase-locked loop (see Methods). We then record the frequency and amplitude of this oscillator as a function of time, using an averaging time of ~160 ms. Figure 4b plots the distributions of the measured frequency and amplitude in the conventional nonlinear case as the purple circles for different driving amplitudes of the feedback loop. The phase delay $\Delta\varphi$ between the oscillations and the drive of the phase-locked loop is adjusted for maximum vibration amplitude, so that the center of the distributions coincides with the peaks of the corresponding resonance response curves measured when the driving frequency is swept with the feedback turned off (blue circles). The width of the distributions along the frequency axis, which is related to the standard deviation $\sigma_\omega$ of the frequency, display a non-monotonic dependence on the amplitude of the active drive. As shown in Fig. 4b for a monotonic transformed backbone line, a stronger drive initially improves the phase stability for small oscillation amplitudes. The frequency standard deviation $\sigma_\omega$ decreases when the driving amplitude $F_{ac}$ is increased from 9.5 pN to 63.35 pN. However, at large vibration amplitudes, the nonlinear effects convert amplitude noise into phase noise, leading to an increase in $\sigma_\omega$ when the driving amplitude is increased to 158.4 pN. The blue curve in Fig. 4d plots $\sigma_\omega$ versus the driving amplitude of the feedback loop for the case when the transformed backbone line is monotonic, showing an optimal driving amplitude that yields the minimum standard deviation: $\sigma_\omega^{(min)} \approx 0.106$ rad s$^{-1}$ at $F_{ac} \approx 63$ pN. Here, the data is taken in a separate cool down compared to Figs. 1-3. The noise intensity $D_w$ is fixed at 6800 pN$^2$ Hz$^{-1}$ for self-sustained oscillations and set to zero for the resonance response curves without feedback.

The frequency stability can be significantly improved when $V_{dc}$ is adjusted to yield zero-dispersion behavior. Figure 4c shows measurements for the case of a non-monotonic backbone line using four driving amplitudes (22.2 pN, 71.3 pN, 182 pN and 269 pN). For force vibrations without feedback, the driving amplitude of 182 pN (yielding the third resonance curve from bottom) marks the transition of the resonance curves from bending towards high frequency to the mixed behavior. For self-sustained oscillations with feedback, the distribution of measured frequencies (plotted in purple) shows a remarkable narrowing for this driving amplitude as the system approaches the zero dispersion regime, where the eigenfrequency of the resonator is locally independent of the vibration amplitude. In Fig. 4d, the dependence of $\sigma_\omega$ on driving amplitude for this case is plotted in red, showing that the initial drop of $\sigma_\omega$ with driving amplitude largely coincides with the blue curve for a nonlinear resonator with a monotonic transformed backbone line. As the driving amplitude continues to increase beyond 60 pN, the two oscillators behave differently. Instead of a sharp increase with the driving amplitude, $\sigma_\omega$ for the resonator with non-monotonic backbone exhibits only a slight increase followed by a gentle drop. The drop becomes steeper as the driving amplitude further increases. $\sigma_\omega$ eventually attains the absolute minimum $\sigma_{\omega,\text{zd}}^{(\min)} \approx 0.034$ rad s$^{-1}$ at driving amplitude of 182 pN (same as the driving amplitude for the second highest resonance response curve in Fig. 4c), beyond which $\sigma_\omega$ increases sharply. Remarkably, the minimal $\sigma_\omega$ for the zero-dispersion regime at 182 pN is about 3 times smaller compared to the minimal $\sigma_\omega$ in the conventional regime. Further improvements appear feasible if the zero-dispersion energy can be pushed higher.

In analyzing the frequency stability of oscillators and clocks, the Allan deviation is often used to eliminate systematic errors such as long term frequency drifts. For the results shown in Fig. 4d, the frequency of the phase locked loop is recorded for only 120 s for each driving amplitude. Over such a short duration, the effect of frequency drift is negligible in our system. In Supplementary Note 4, we show that the improvement of the minimal Allan deviation in the zero dispersion regime over its counterpart in the conventional regime is about 3.6 for averaging time of 1 s, close to the improvement in $\sigma_\omega$ in Fig. 4d.

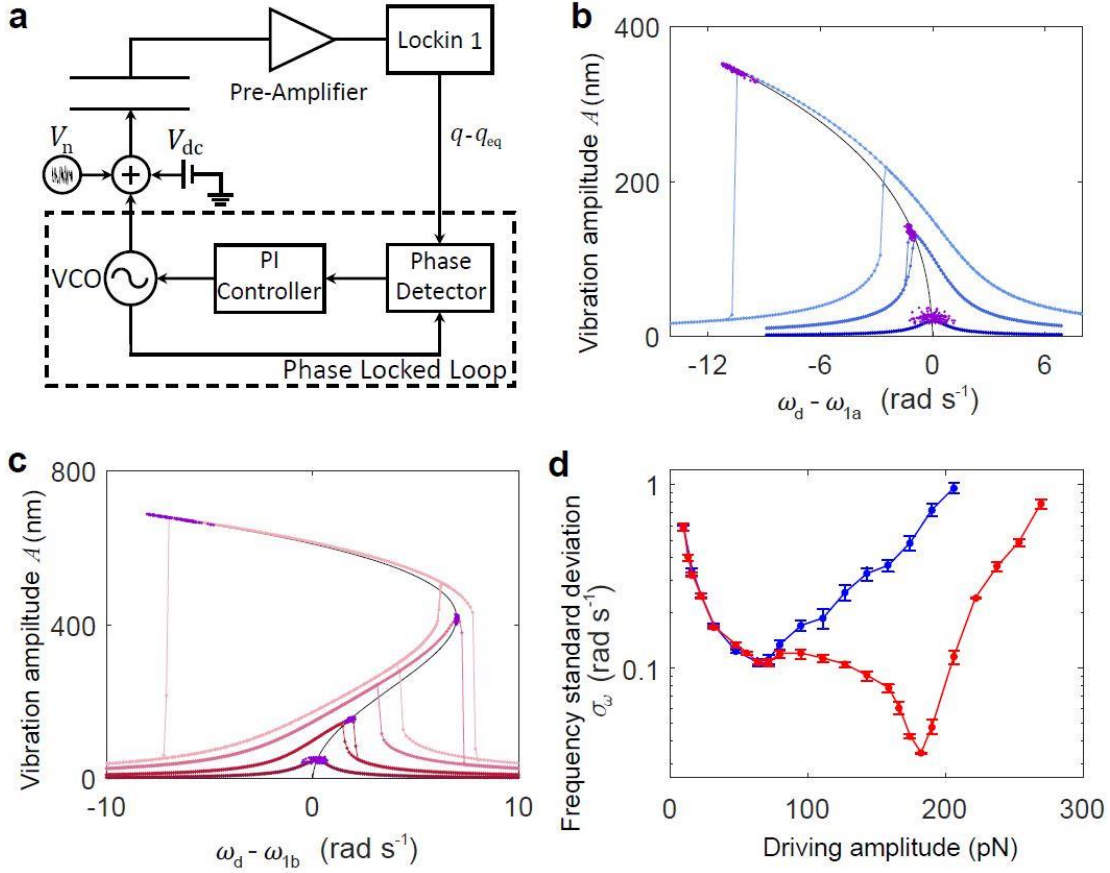

**Fig. 4** Frequency stabilization in the zero–dispersion regime compared to the conventional regime. **a** Schematic for driving the resonator into self-sustained oscillations with a phase-locked loop. **b** Forced and self-sustained oscillations in the conventional nonlinear regime. With no feedback, the forced vibration amplitude as a function of driving frequency is measured at driving amplitudes of (bottom to top) 9.5 pN, 63.35 pN and 158.4 pN. The solid black line represents the backbone line. With the feedback turned on and the phase delay chosen to maximize the self-oscillation amplitude, the measured distributions of the self-sustained oscillation amplitude and frequency are plotted in purple. **c** Similar plot for the zero dispersion regime. From bottom to top, the driving amplitudes are 22.2 pN, 71.3 pN, 182 pN and 269 pN. The stabilization of frequency in the zero-dispersion regime is demonstrated in the third curve at driving amplitude of 182 pN. **d** Standard deviation of the frequency of self-sustained oscillations $\sigma_\omega$ versus driving amplitude for oscillators in the conventional regime (blue) and the zero-dispersion regime (red). The zero-dispersion regime yields the minimum standard deviation $\sigma_\omega^{(min)}$ about a factor of ~ 3 smaller than that of the conventional regime. Error bars represent ±1 s. e.

**Discussion**

Our findings show that the spectral width of micromechanical resonators is determined by the interplay of damping, nonlinearity and additive noise. In a conventional nonlinear oscillator, where the transformed backbone line is monotonic, the fluctuation spectrum peak widens as the noise intensity increases. For a resonator with zero dispersion, where the transformed backbone line is non-monotonic, the spectral peak undergoes narrowing as the noise intensity is increased in some range related to $E_{zd}$, in agreement with theoretical calculations.

For the zero-dispersion resonator, the optimal frequency stability of self-sustained oscillations shows a considerable improvement as compared to conventional nonlinear resonator. One advantage of such an approach to stabilize the frequency in the presence of strong nonlinearity is that only a single vibrational mode is involved. Matching the frequency of other modes[21] is therefore unnecessary, potentially simplifying the operation.

The zero dispersion in our resonator is obtained by tuning the negative electrostatic nonlinearity relative to the positive intrinsic nonlinearity of the springs. Non-monotonicity of a transformed backbone line can also be achieved using other methods, such as the coupling of two nanomechanical resonators[51] or the negative nonlinear friction induced by dynamical backaction from a photon or phonon cavity[52]. Thus, zero-dispersion is a rather generic feature of mechanical resonators and such resonators provide a well-controlled platform to investigate and exploit zero-dispersion phenomena reported in the present work as well as others that have not yet been experimentally observed.

The improvement of frequency stability for resonators driven into self-sustained vibrations using feedback was demonstrated for applied noise that is additive. Such noise plays a role similar to thermal-mechanical noise that leads to both phase and amplitude fluctuations. At small vibration amplitudes, the direct contribution to phase fluctuations dominates. As the amplitude increases, conventional nonlinearity converts amplitude noise into phase noise. The effect of such a conversion is substantially reduced in the zero-dispersion regime with $d\omega/dE = 0$ so that amplitude fluctuations do not affect the phase stability even at large vibration amplitudes. If, instead, the phase noise is dominated by fluctuations in the eigenfrequency of the resonator[13-16], operating the feedback loop in the regime of zero-dispersion does not yield improvement in phase stability because amplitude fluctuations are not involved.

The main effect of the injected noise in our experiment is that it increases the effective temperature of the resonator. If the injected noise is removed, the thermal noise at temperature of 4 K leads to random motion of the plate. In our experiment, such thermal motion is too small to be resolved due to noise in the detection circuit at room temperature. To estimate the effect of zero-dispersion on preventing the conversion of amplitude fluctuations originating from thermal noise into phase fluctuations, we perform numerical simulations assuming that no additional noise is introduced by the feedback loop and the detection circuit. The optimal $\sigma_\omega$ in the zero-dispersion regime is found to be smaller than the conventional nonlinear regime by a factor of 4, comparable to the case measured in our experiment with injected noise.

Apart from the applied noise and thermal-mechanical noise, the feedback circuit also contributes to noise in the phase of oscillations. In particular, we consider the effects from the output of the voltage-controlled oscillator (VCO) that is used to drive the mechanical resonator. Fluctuations in the VCO output amplitude leads to amplitude noise in vibrations, but the phase dynamics are not directly affected if the resonator is linear. However, when the vibration amplitude is large, nonlinearity becomes strong and such amplitude noise could be converted into phase fluctuations. Designing and operating the resonator in the zero-dispersion regime reduces such conversion and is therefore expected to yield improved frequency stability, similar to the case of additive noise reported here. The experimental demonstration of such improvement, however, is outside the scope of this paper.

For the non-monotonic transformed backbone lines, the optimal phase stability for self-sustained oscillations is achieved at the vibration amplitude $A_{zd}$ at which $d\omega/dE = 0$. By eliminating the extra contributions to phase noise converted from amplitude noise, the phase stability becomes comparable to an oscillator undergoing purely linear vibrations at amplitude $A_{zd}$. It is well-known that for linear resonators, the frequency standard deviation for self-sustained vibrations is inversely proportional to the vibration amplitude[53]. Therefore, the optimal frequency standard deviation for an oscillator in the zero dispersion regime is expected to follow a similar dependence on $A_{zd}$, i.e. $\sigma_{\omega,zd}^{(min)} \propto 1/A_{zd}$. Ideally, the phase stability can be improved by choosing $A_{zd}$ to be as large as possible. The upper limit in our experiment is given by the gap $g$ between the movable plate and the fixed electrode. In practice, operating close to this upper limit could lead to the plate touching the electrode when occasional large fluctuations occur. In Figs. 3 and 4, $A_{zd}$ is ~ 400 nm, about a factor of 5 smaller than $g$. Assuming that $A_{zd}$ can be increased to 80% of the upper limit, the ratio of the optimal frequency standard deviation in the zero dispersion regime to that in the conventional regime could reach ~ 12, a factor of ~4 larger than in Fig. 4d. Demonstrating such best case scenario in our device, however, is not the main goal of this paper. Instead, we choose $A_{zd}$ to be smaller than $g$ to show the general features of zero dispersion phenomena. For example, to reveal the local minimum of the spectral width as a function of the noise intensity in Fig. 3d, it is necessary to go beyond the zero-dispersion regime using vibration amplitude that approaches the maximum displacement. Promising future research directions for phase stabilization using zero dispersion include new approaches to yield large $A_{zd}$, as well as the effect on the phase stability due to other phenomena that emerges when the vibration amplitude becomes large, such as nonlinear friction.

Finally, we note that zero-dispersion phenomena often become more prominent when the ratio $(\omega_{zd} - \omega_l)/\Gamma$ increases. In the current experiment, this ratio is ~ 10. If the ratio can be significantly increased in future designs, a wealth of zero-dispersion phenomena is expected to occur. Examples include the appearance of extremely sharp peaks in fluctuation spectra[29, 32, 38] and the onset of deterministic chaos at extraordinarily low periodic drive amplitudes[37, 38, 54], to name just a few. Further studies of these and other zero-dispersion phenomena may not only be of fundamental interest, but could also lead to new methods of signal detection and other applications.

During the preparation of this manuscript, the authors learned[55] that the theory for reduction of phase noise in self-sustained oscillations exploiting the zero-dispersion property is being developed by Miller *et al*[56, 57]. These results could allow for further improvements in phase stability using zero-dispersion phenomena.

## Methods

**Detection scheme.** Displacement of the top plate is inferred from the capacitance change $\Delta C$ between the top plate and the fixed electrode underneath. As shown in Fig. 1c, two ac voltages $V_{c1}$ and $V_{c2}$ at the same frequency $f_c = \omega_c/2\pi$ (4 MHz), with comparable amplitude (300 mV) and opposite phases, are applied to the bottom electrode and a standard capacitor, respectively. The other plate of the standard capacitor and the top plate of the resonator are electrically connected. Displacement of the top plate leads to a change in the capacitance between the top plate and the fixed electrode. Therefore, motion of the top plate modulates the amplitude of ac signal at the carrier frequency on the top plate. This ac signal is measured with a lock-in amplifier referenced to $f_c$. Changes in the lock-in output are proportional to the deviation of the displacement $q$ from the equilibrium position $q_{eq}$.

**Noise generation.** The noise voltage is generated from the Johnson noise of a 50 Ω resistor at room temperature. After the Johnson noise is amplified and passed through a band-pass filter with center frequency $f_{center}$ and bandwidth $f_{bd}$, it is mixed with a noise carrier voltage at frequency 50 kHz, creating two sidebands centered at 50 kHz $\pm f_{center}$. In the experiments on the fluctuation spectra measurement, $f_{center} = 23034$ Hz and $f_{bd} = 741$ Hz for the conventional nonlinear regime ($V_{dc} = -2.3$ V), while $f_{center} = 23143$ Hz and $f_{bd} = 298$ Hz for the zero-dispersion regime ($V_{dc} = -1.51$ V). In the experiments on the frequency stabilization, $f_{center} = 23237$ Hz and $f_{bd} = 748$ Hz for the conventional nonlinear regime ($V_{dc} = -1.863$ V), while $f_{center} = 23271$ Hz and $f_{bd} = 749$ Hz for the zero-dispersion regime ($V_{dc}$ = -1.59 V). The eigenfrequency of the resonator lies within the upper sideband. Its bandwidth is much larger than the width of the spectral peaks discussed in the main text.

**Measurement of the spectral density of fluctuations.** When we measure the spectral density of fluctuations, the periodic drive is turned off. $V_n(t)$ produces a random force that plays a similar role as thermo-mechanical noise. The first lock-in amplifier produces an output that is proportional to $q$. A second lock-in amplifier, referenced at frequency $\omega_l/2\pi$ and with a bandwidth of 100 Hz yields amplitudes of oscillations $X(t)$ and $Y(t)$ that are in phase and out of phase with the reference:

$$q(t) = X(t)\cos(\omega_l t) + Y(t)\sin(\omega_l t). \tag{18}$$

The spectral density of fluctuations is calculated then as follows:

$$\tilde{Q}(\omega) = \frac{1}{N}\sum_\tau \sum_t \left\{ [X(t+\tau) + iY(t+\tau)][X(t) - iY(t)]e^{-i(\omega-\omega_d)\tau} \right\}, \tag{19}$$

where the discretization number $N$ is 4000 and 4044 in the conventional and zero-dispersion cases respectively. The discretization step (identical for $\tau$ and $t$) is 4.44 ms and 8.93 ms in the conventional and zero-dispersion cases respectively.

**Oscillations sustained by active feedback.** Lock-in amplifier 1 produces a signal $V_{out1}(t)$ that is proportional to the deviation of the plate displacement $q(t)$ from the equilibrium position $q_{eq}$. Self-sustained oscillations are maintained by feedback using the phase-locked loop of lock-in amplifier 2. Specifically, $V_{out1}(t)$ is fed into the reference of lock-in amplifier 2. The phase-locked loop adjusts the frequency $\omega_d$ of the periodic drive to maintain a fixed, controllable phase delay $\Delta\varphi$ between $V_{out1}(t)$ and the periodic drive. $\Delta\varphi$

is adjusted to maximize the oscillation amplitude. Data in Fig. 4 are taken in a separate cool down from Figs. 1 to 3 so that the device parameters are slightly changed. For the conventional regime (Fig. 4b), $V_{dc}$ is chosen to be -1.863 V, giving $\omega_l = \omega_{la} \equiv 460180.7$ rad s$^{-1}$ and a transformed backbone line with $\kappa = -6.19 \times 10^{13}$ rad s$^{-1}$ m$^{-2}$ and $\eta = -2.23 \times 10^{26}$ rad s$^{-1}$ m$^{-4}$. For the zero-dispersion regime (Fig. 4c), $V_{dc}$ is -1.59 V, giving $\omega_l = \omega_{lb} \equiv 460376.2$ rad s$^{-1}$ and a transformed backbone line with $\kappa = 6.12 \times 10^{13}$ rad s$^{-1}$ m$^{-2}$ and $\eta = -1.39 \times 10^{26}$ rad s$^{-1}$ m$^{-4}$. $\kappa$ for the two different $V_{dc}$ have almost the same magnitude but opposite sign.

**Data availability**

The data that support the findings of this study are available from the corresponding author on request.

**Acknowledgements**

L.H. and H.B.C. are supported by Research Grant Council of HKSAR, China (project No. 16303215). K.N. is supported by NSF No. DMR- 0645448. S.M.S. acknowledges the support by Volkswagen Foundation (Grant No. 90418), the support of his visits to Pisa by Pisa University and to University of Warwick by the Institute of Advanced Studies and School of Engineering of University of Warwick, and the access to the e-library of Lancaster University provided by his Honorary Visiting Researcher position there.

**Author Contributions**

H.B.C. conceived the idea of the work and designed the experiments. S.M.S., I.A.K. and R.M. developed the theory. L.H. and K.N. performed the experiments and analyzed the data. L.H., H.B.C. and S.M.S. co-wrote the paper.

**Competing Interest Statement**

The authors declare no competing financial interests.

# Frequency stabilization and noise-induced spectral narrowing in resonators with zero dispersion

Huang *et al.*

Supplementary Note 1. Theory of frequency of eigenoscillation vs. its amplitude and theoretical calculation of resonance response curves

Eigenoscillation is defined as an oscillation in an idealized conservative system. It involves theoretical approximation that neglects dissipative and time-dependent forces. In relation to the system defined by Eqs. (1)-(4), it means that we should neglect in the equation of motion (1) the friction $-2\Gamma\dot{q}$ and average an explicit dependence of other forces on time, if any. The spring restoring force does not explicitly depend on time and therefore the averaging has no effects on it. The electric force may explicitly depend on time and thereby, for $V$ in Eq. (4), we replace $V^2$ in Eq. (3) for the electric force by its average over time $\overline{V^2}$:

$$\overline{V^2} = V_{dc}^2 + V_{ac}^2/2 + \langle V_n^2 \rangle + V_c^2/2 \qquad (1)$$

The non-resonant (high-frequency) ac voltage $V_{c1}(t) \equiv V_c \cos(\omega_c t)$ is present in all our experiments, so that $V_c^2/2$ is to be kept in $\overline{V^2}$ in all cases. As for $V_{ac}^2/2$, it is either exactly equal to zero (in the experiments on fluctuation spectra) or being much smaller than $V_c^2/2$. Therefore we neglect it in all cases. Finally, in those experiments where noise is present, the average of its square $\langle V_n^2 \rangle$ is much smaller than $V_{dc}^2$ (and even $V_c^2/2$). Nevertheless we do take it into account in $\overline{V^2}$ in order to further improve the agreement with the experiment (see Supplementary Note 3 and Figs. 3a and 3b in the main text). In the present Supplementary Note however, we assume for the sake of simplicity of notations that noise is absent or negligible i.e.

$$\overline{V^2} = \tilde{V}_{dc}^2 \equiv V_{dc}^2 + V_c^2/2. \qquad (2)$$

Next, we discuss the calculation of the frequency of eigenoscillation $\omega$ vs. its amplitude $a$, assuming that $\tilde{V}_{dc}^2$ has a given value from the range defined in Eq. (6) while $a$ is sufficiently small for the deviation of $\omega(a)$ from its zero-amplitude limit $\omega(a\to 0) \equiv \omega_1$ to be small i.e. $|\omega(a) - \omega_1| \ll \omega_1$. After the omission of friction in the equation of motion and the replacement of $V^2$ by $\overline{V^2} = \tilde{V}_{dc}^2$, we transform from $q$ to $x = q - q_{eq}$ where $q_{eq} \approx \tilde{v}g$ is the equilibrium position, expand the force into the Taylor series in powers of $x$ and, allowing for the smallness of $\tilde{v}$ in the relevant range of $\tilde{V}_{dc}^2$, neglect in coefficients of the Taylor series terms proportional to $\tilde{v}^n$ with $n \geq 2$. As can be shown (cf. [1,2]), it is sufficient to keep in the Taylor series only terms up to the 5th power if we are interested only in the quartic approximation for $\omega(a)$ (see Eq. (11)) while the latter, in turn, is sufficient for our case. So, we end up with the following equation of motion:

$$\ddot{x} = -\sum_{n=1}^{5} \alpha_n x^n, \tag{3}$$

where $\alpha_n$ are given in Eqs. (8)-(10) of the main text and in the paragraph following Eq. (10). Next, we apply to this system the method of successive approximations described in Refs. [1,2]. Namely, we seek the solution of the equation of motion in the form of series $\sum_{i=1}^{\infty} x^{(i)}$ where $x^{(i)}$ with a given $i$ is proportional to $a^i$ while

$$x^{(1)} = a\cos(\omega t), \tag{4}$$

where $\omega \equiv \omega(a)$ is the exact frequency of eigenoscillation the first harmonic of which has the amplitude $a$. In turn, $\omega(a)$ is sought in the form of series $\sum_{i=0}^{\infty} \omega^{(i)}$ where $\omega^{(i)}$ with a given $i$ is proportional to $a^i$ while

$$\omega^{(0)} = \sqrt{\alpha_1}. \tag{5}$$

The coefficient $\omega^{(0)}$ (Supplementary Equation 5) immediately gives us the formula for $\omega_1$ in the approximation of $\omega(a)$ by Eq. (11). Allowing for Eq. (8) for $\alpha_1$, we obtain from it Eq. (12).

In order to find the next-order terms in the series for $x$ and $\omega$, i.e. $x^{(2)}$ and $\omega^{(1)}$, we first present the equation of motion (Supplementary Equation 3) in the following equivalent form

$$\frac{\left(\omega^{(0)}\right)^2}{\omega^2}\ddot{x} + \left(\omega^{(0)}\right)^2 x = -\sum_{n=2}^{5} \alpha_n x^n - \left(1 - \frac{\left(\omega^{(0)}\right)^2}{\omega^2}\right)\ddot{x}, \qquad (6)$$

then substitute into it the series for $x$, and neglect in both sides of the equation small terms proportional to $a^i$ with $i \geq 3$. Terms proportional to $a$ cancel each other owing to the proper choice of $x^{(1)}$ and $\omega^{(0)}$, so that we end up with the closed equation for terms of the second order in $a$. Using the identity $(\cos(\omega t))^2 = (1 + \cos(2\omega t))/2$, we present this equation as follows:

$$\ddot{x}^{(2)} + \left(\omega^{(0)}\right)^2 x^{(2)} = -\frac{\alpha_2 a^2}{2} - \frac{\alpha_2 a^2}{2}\cos(2\omega t) - 2a\omega^{(0)}\omega^{(1)}\cos(\omega t). \qquad (7)$$

The above equation has a simple interpretation: it is an equation of motion of the linear (harmonic) oscillator with the eigenfrequency $\omega^{(0)}$ (left-hand side of the equation) driven by forces of two distinctly different types in the present context. The first type is represented by the first and second terms on the right-hand side: they are distinctly non-resonant. The second type is represented by the last term on the right-hand side: it is almost resonant. But the constrained vibrations under action of an almost resonant force would have a large amplitude, which would contradict to the original assumption that an amplitude of each successive correction to the solution is small. In order to avoid such a contradiction, the value of $\omega^{(1)}$ should be chosen in such a way that the amplitude of the "almost resonant force" would turn into zero. The latter takes place only if we impose the condition

$$\omega^{(1)} = 0. \qquad (8)$$

This value of $\omega^{(1)}$ provides the deletion of the almost resonant force from Supplementary Equation 7. Solving the inhomogeneous linear differential equation (Supplementary Equation 7) with the remaining inhomogeneous part in a usual way, we have

$$x^{(2)} = -\frac{\alpha_2 a^2}{2(\omega^{(0)})^2} + \frac{\alpha_2 a^2}{6(\omega^{(0)})^2}\cos(2\omega t). \tag{9}$$

In order to find the next-order terms in the series for $x$ and $\omega$, i.e. $x^{(3)}$ and $\omega^{(2)}$, we perform the analysis analogously to the previous step: substitute the series in the equation of motion (Supplementary Equation 6) and neglect terms which are proportional to $a^i$ with $i \geq 4$ while terms proportional to $a$ cancel each other owing to the proper choice of $x^{(1)}$ and $\omega^{(0)}$ and terms proportional to $a^2$ cancel each other owing to the proper choice of $x^{(2)}$ and $\omega^{(1)}$, so that we end up with the closed equation for terms of the third order in $a$. Similar to the previous step, the resulting differential equation for $x^{(3)}$ represents the equation of motion of the linear oscillator driven both by a distinctly non-resonant force and by an almost resonant one:

$$\ddot{x}^{(3)} + \left(\omega^{(0)}\right)^2 x^{(3)} = -a^3\left(\frac{\alpha_3}{4} + \frac{\alpha_2^2}{6(\omega^{(0)})^2}\right)\cos(3\omega t) + a\left(2\omega^{(0)}\omega^{(2)} + \frac{5a^2\alpha_2^2}{6(\omega^{(0)})^2} - \frac{3a^2\alpha_3}{4}\right)\cos(\omega t). \tag{10}$$

Then we again apply one of the key ideas of the method: we find $\omega^{(2)}$ from the condition that the amplitude of the almost resonant force should turn into zero. The result is the following:

$$\omega^{(2)} = a^2\left(\frac{3\alpha_3}{8\omega^{(0)}} - \frac{5\alpha_2^2}{12(\omega^{(0)})^3}\right). \tag{11}$$

Solving then the resulting equation (Supplementary Equation 10) in a usual manner, we easily find $x^{(3)}$.

One can continue such iterative process to any order. It may be worth noting however that, as the order of an approximation further increases, calculations become very cumbersome: even the ultimate expressions are quite unwieldy. Thus, $\omega^{(4)}$ is given by the following formula (note that $\omega^{(3)}$, like any other odd-order correction, is equal to zero):

$$\omega^{(4)} = a^4 \frac{1}{16\,\omega^{(0)}} \left( 5\alpha_5 - \frac{15\alpha_3^2}{16(\omega^{(0)})^2} - \frac{14\alpha_2\alpha_4}{(\omega^{(0)})^2} + \frac{22\alpha_3\alpha_2^2}{3(\omega^{(0)})^4} - \frac{49\alpha_2^4}{36(\omega^{(0)})^6} \right). \tag{12}$$

In our case however, the situation greatly simplifies due to the smallness of the parameter $\tilde{v}$. In particular, the even coefficients $\alpha_{2i}$ are proportional to $\tilde{v}$ and it is seen from Supplementary Equations 11 and 12 that all terms which include one or more even coefficients $\alpha_{2i}$ with $i \geq 1$ are proportional to $\tilde{v}^n$ with $n \geq 2$ and therefore may be neglected in comparison with other terms.

Consider first Supplementary Equation 11 for $\omega^{(2)}$. Dividing it by $a^2$, neglecting terms including even coefficients $\alpha_{2i}$, allowing for the formula (Supplementary Equation 5) for $\omega^{(0)}$ (together with the first-order approximation (8) for $\alpha_1$) and using the first-order approximation (9) for $\alpha_3$, we obtain Eq. (13) for the coefficient $\kappa$ in the quadratic term of the quartic approximation (11) for $\omega(a)$.

The situation with Supplementary Equation 12 is more subtle. Dividing it by $a^4$, neglecting terms including the even coefficients ($\alpha_2$ or $\alpha_4$), using the zero-order approximation (8) for $\alpha_1$ and first-order approximations (9) and (10) for $\alpha_3$ and $\alpha_5$ respectively, and taking into account the definition of the nonlinearity length scale $L_\text{n} \equiv$ $\omega_s/\sqrt{\beta_s}$, we obtain the following expression for the coefficient $\eta$ in the quartic term of the quartic approximation (11) for $\omega(a)$:

$$\eta = -\frac{15\omega_s\tilde{v}}{8g^4}\left\{1 - \frac{8\mu_s L_n^2}{3\beta_s}\frac{[g/(2L_n)]^2}{\tilde{v}}\left(\frac{g}{2L_n}\right)^2 + \frac{1}{2}\tilde{v}\left(\frac{[g/(2L_n)]^2}{\tilde{v}} - 1\right)^2\right\}. \tag{13}$$

The third term in the curly parentheses is $\lesssim \tilde{v}$ in the relevant range indicated in Eq. (6), where $\tilde{v}/[g/(2L_n)]^2 \sim 1$, and therefore it can be neglected as compared to the first term 1. Furthermore, characteristic scales of spring nonlinearities of close orders are typically close to each other, in particular this concerns nonlinearities of the third and fifth orders and, therefore, $\mu_s L_n^2/\beta_s \sim 1$. Hence, for the relevant range indicated in Eq. (6), the second term in the parentheses is of the order of $[g/(2L_n)]^2 \ll 1$ so that it can also be neglected. Thus, Supplementary Equation 13 for $\eta$ reduces to Eq. (14), meaning that $\eta$ is strongly dominated by the electrostatic contribution and is distinctly negative, which nicely conforms to the experiment.

As for $\kappa$, the ratio of the contributions from the electric and spring forces is represented by the second term in the parentheses of Eq. (13). This term is a ratio of two small parameters. The most interesting (in the zero-dispersion context) part of the range of the dc voltage indicated in Eq. (6) corresponds to the case when $\tilde{v}$ is barely exceeded by $[g/(2L_n)]^2$ (in other words, when the spring contribution weakly dominates over the electrostatic one), that yields a small but positive $\kappa$.

Since $\kappa$ is positive while $\eta$ is negative, $\omega(a)$ (11) possesses a local maximum and, moreover, as the excess of 1 over the ratio $\tilde{v}/[g/(2L_n)]^2$ is small, the maximum lies in the range of $a$ being much smaller than $g$, where the quartic approximation (11) adequately describes the true $\omega(a)$, that proves the self-consistency of the theory.

If $\tilde{v}$ is much smaller than $[g/(2L_n)]^2$ (i.e. if $\tilde{V}_{dc}^2 \ll V_{zd}^2$), then the above arguments in favor of the domination of the electrostatic contribution over the spring contribution are no longer valid and, in fact, it is vice versa for sufficiently small values of $\tilde{v}$: the spring contribution dominates i.e. $\eta \approx \eta_s$, thus being positive. However it does not mean that the resonator does not possess the zero-dispersion property in this case i.e. that $\omega(a)$ does not possess a local maximum: it does, which can be shown by a different method (which will be presented by us elsewhere), but the maximum occurs at much larger values of $a$, namely the relevant range lies close to $g$ and the quartic approximation (11) is evidently insufficient for an adequate description of $\omega(a)$.

Thus we conclude that the value $V_{zd}^2 \equiv \frac{\beta_s m g^5}{2\varepsilon S}$, corresponding to the zero value of $\kappa$, represents the boundary between zero-dispersion and conventional nonlinear regimes in terms of $\tilde{V}_{dc}^2$. It may be worth noting that it does not depend on the elasticity of the springs i.e. on the Hooke's coefficient.

Consider now the resonator subject to a weak resonant ac driving $F_{ac}\cos(\omega_d t)$. The theoretical approximation of the resonance curves $A(\omega_d)$ is determined by the equation[1,2]

$$4m^2\omega_1^2 A^2\{[\omega(a=A) - \omega_d]^2 + \Gamma^2\} = F_{ac}^2, \tag{14}$$

where the eigenoscillation amplitude *a* in $\omega(a)$ is replaced with the amplitude of constrained vibrations *A*.

In experiments, $\omega(a=A)$ is measured by recording the resonance curves at many values of the driving amplitude. By identifying the peak of the curves, we obtain the approximate backbone line $A_{peak}(\omega_{peak})$[1]. For any given value of $A_{peak}$, the reversed function $\omega_{peak}(A_{peak})$ is approximately equal to the frequency of eigenoscillation with the amplitude equal to the given $A_{\text{peak}}$, i.e. $\omega(a) \approx \omega_{\text{peak}}(A_{\text{peak}} = a)$. The transformed backbone line (e.g. in Fig. 2b and Fig. 2d) of $\omega_{\text{peak}}$ vs. $A_{\text{peak}}^2$ well approximates the eigenfrequency vs. the scaled energy:

$$\omega(E) \approx \omega(a^2 = E/\{m\omega_1^2/2\}) \approx \omega_{\text{peak}}(A_{\text{peak}}^2 = E/\{m\omega_1^2/2\}). \tag{15}$$

Furthermore, the measured transformed backbone line is well fitted by the parabolic function both in the zero-dispersion and conventional regimes (Figs. 2b and 2d respectively), that allows us to find the corresponding coefficients $\kappa$ and $\eta$ in the approximation of $\omega(a)$ (11). This approximation of $\omega(a)$ was substituted then in Supplementary Equation 14 in order to find the theoretical approximations of the resonance curves $A(\omega_d)$ for given values of the dc voltage $V_{\text{dc}}$ and the driving amplitude $F_{\text{ac}}$: see thin solid lines in Figs. 2a and 2c. The agreement with the experimental lines is satisfactory.

Supplementary Note 2. Effect of the quadratic noise and the white-noise approximation.

There are just few requirements to the form of the voltage noise component $V_n(t)$ which are crucial for the experiment: its spectrum (spectral power density) should represent a narrow peak with the maximum close to $\omega_1$ and the width greatly exceeding the width of the frequency range relevant to the fluctuation spectrum peaks in a vicinity of $\omega_1$. On the other hand the width of the $V_n$ spectrum peak should be sufficiently small for the quadratic noise to be negligible (otherwise such noise may smear zero-dispersion effects). For the sake of brevity and clarity, we consider below only the form of the noise which was used in our experiments (still keeping some extent of generality where possible) while a generalization is straightforward.

$V_n$ used in the experiments can be presented in the following form:

$$V_n(t) = N_c(t)\cos(\omega_1 t) + N_s(t)\sin(\omega_1 t),$$

(16)

where $N_c(t)$ and $N_s(t)$ are identical independent noise sources with the zero average and a monotonously decaying correlation function,

$$\langle N_c(t) \rangle = \langle N_s(t) \rangle = 0, \tag{17}$$

$$\langle N_c(t)N_s(t') \rangle = \langle N_s(t)N_c(t') \rangle = 0, \tag{18}$$

$$\langle N_c(t)N_c(t') \rangle = \langle N_s(t)N_s(t') \rangle = 2I_n d(t-t'), \tag{19}$$

where $t-t' \geq 0$ and $d(\tau)$ is a monotonously and slowly (as compared to the period of eigenoscillation $2\pi/\omega_1$) decaying function normalized similarly to the $\delta$–function:

$$\int_0^\infty d\tau\, d(\tau) = \frac{1}{2}. \tag{20}$$

Using Supplementary Equations 16 to 19, we can easily derive formulas both for the average and for the correlation function of $V_n$:

$$\langle V_n(t) \rangle = 0, \qquad \langle V_n(t)V_n(t') \rangle = 2I_n d(t-t')\cos[\omega_1(t-t')]. \tag{21}$$

It may be expected that $d(\tau)$ is characterized with a single decay time-scale $(2d(0))^{-1}$ which may also be interpreted as a correlation time-scale $t_{cor} \equiv (2d(0))^{-1}$. Measurements confirm that $V_n(t)$ in our experiment satisfies the above requirement. Indeed, the smoothed spectrum has a Gaussian form (Supplementary Figure 1):

$$Q^{(V_n)}(\Omega) \equiv \frac{1}{\pi} \text{Re}\left[\int_0^\infty dt\, \langle V_n(t)V_n(0) \rangle \exp(-i\Omega t)\right] \approx \frac{I_n}{2\pi} \exp\left\{-\frac{1}{2}\left(\frac{\Omega-\omega_1}{\pi f_{bd}/\sqrt{2\ln(2)}}\right)^2\right\}, \tag{22}$$

and therefore $d(\tau)$ is Gaussian as well, namely:

$$d(\tau) = \frac{1}{2t_{cor}} \exp\left\{-\frac{\pi}{4}\left(\frac{\tau}{t_{cor}}\right)^2\right\}, \qquad t_{cor} \equiv \frac{\sqrt{\ln(2)/\pi}}{f_{bd}}. \qquad (23)$$

The noise intensity $I_n$, which may also be interpreted as a product of $2\pi$ and the maximum of the noise power spectrum $Q_{max}^{(V_n)} \approx Q^{(V_n)}(\omega_1)$ (Supplementary Figure 1), is small in the sense that $\langle N_c^2 \rangle = \langle N_s^2 \rangle \equiv \langle N^2 \rangle \ll V_{dc}^2$ so that the quadratic noise plays a negligible role. Allowing for Supplementary Equation 19 and the identity $t_{cor} \equiv (2d(0))^{-1}$, this strong inequality is equivalent to a satisfaction of the following condition:

$$\lambda \equiv \frac{I_n}{t_{cor} V_{dc}^2} \ll 1. \qquad (24)$$

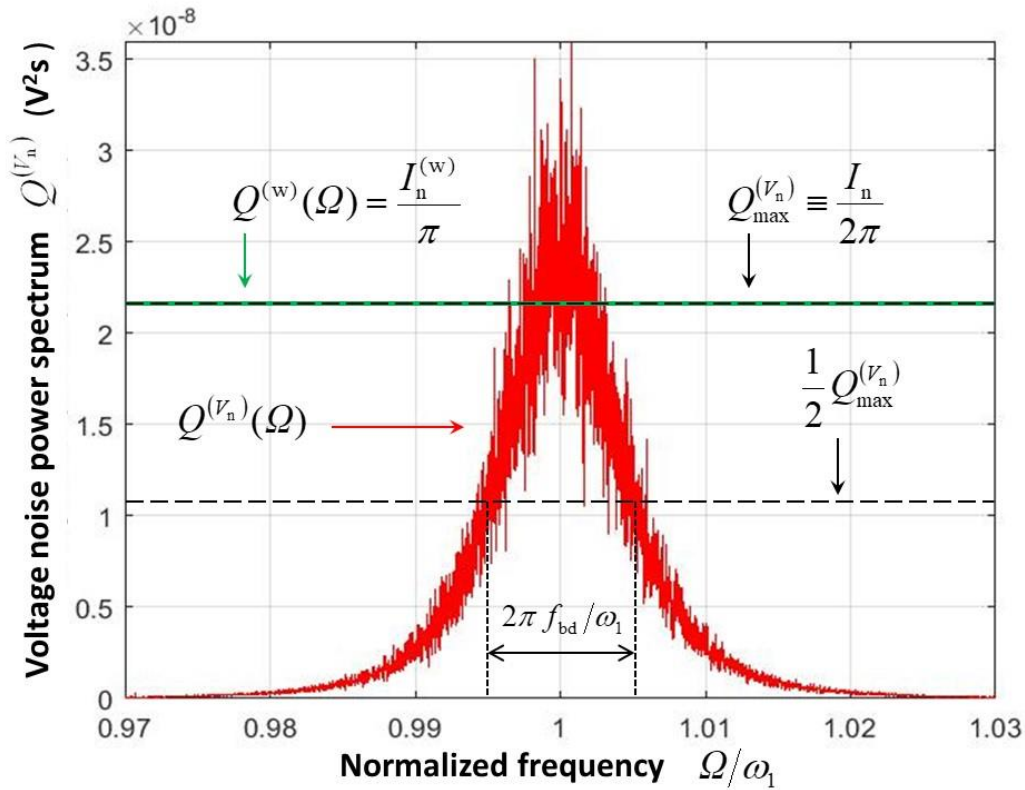

**Supplementary Figure 1** An example of the voltage noise power spectrum. Example of the power spectrum of noise component of voltage $Q^{(V_n)}(\Omega)$ (jagged solid red line): this particular noise acted on the resonator in the conventional regime ($V_{dc} = -2.3$ V) where the noise intensity $D_W = 4470$ pN² Hz$^{-1}$ (see Supplementary Equation 38 below). Fluctuations are due to a limited number of data. Maximum of the smoothed over fluctuations spectrum and its half-level are indicated by the corresponding labels with arrows and by the horizontal dashed lines. The intersection of the lower of the dashed lines with the smoothed spectrum defines the range of the "bandwidth" of the spectrum. The higher dashed line (indicating the maximum) coincides with the thick solid green line showing the white noise spectrum $Q^{(w)}(\Omega)$ of the intensity twice smaller than that of the original noise: its action on the system is almost the same as that by the original noise.

As concerns $t_{cor}$, on the one hand, it greatly exceeds the period of natural oscillation, that allows us in particular to satisfy the condition (Supplementary Equation 24) in the relevant range of noise intensity but, on the other hand, it is much smaller than a reciprocal of the characteristic half-width of any of distinct spectral peaks in the spectrum of fluctuations of the system (1)-(4) (for $V_d \equiv 0$). As it is seen from the consideration below, such smallness of $t_{cor}$ allows us to consider the noise in the context of the fluctuation spectrum as white.

Let us substitute $V$ [Eq.(4)] with $V_d \equiv 0$ and $V_n$ (Supplementary Equation 16) into the equation of motion given by Eqs. (1)-(3), neglect the high-frequency terms (similar to the analysis of the ac-driving), keep only those of noise-induced terms which have the lowest order of the smallness parameter $\lambda$ (the latter procedure includes in particular neglecting the "quadratic" noise term proportional to $\left(N_c^2+N_s^2\right)/2-\langle N^2\rangle$, the intensity of which is of the order of $\lambda$ as compared to that of the linear noise term $F_n/m$ explicated below), and take into account the smallness of the zero-dispersion energy $E_{zd}\equiv m\omega_1^2 q_{zd}^2/2$ as compared to the characteristic energy $m\omega_1^2 g^2/2$ (which allows us to neglect a deviation of $(1-q/g)^2$ from unity in the correlation function of the linear noise term $F_n/m$). The result can be presented in the following form:

$$\ddot{q} = -2\Gamma\dot{q} + \frac{1}{m}\{F_s(q) + F_e(q, V^2 \to \tilde{V}_{dc}^2 + \langle N^2 \rangle)\} + \frac{1}{m}F_n, \qquad (25)$$

where the expression in the curly parentheses may be considered as a generalized potential force $dU^{(n)}(q)/dq$ which just slightly differs from that in the absence of noise,

$$\begin{aligned}
U^{(n)}(q) &\equiv U^{(n)}(q,\lambda) \approx U_s(q) + U_e(q)(1+\lambda) \;, \\
U_s(q) &\equiv \frac{1}{2}m\omega_s^2 q^2 + \frac{1}{4}m\beta_s q^4 + \frac{1}{6}m\mu_s q^6 + \ldots, \\
U_e(q) &\equiv -\frac{\varepsilon S \tilde{V}_{dc}^2}{2(g-q)},
\end{aligned} \qquad (26)$$

and $F_n$, for which we introduce the notion "linear noise force", reads as

$$F_n \equiv F_e\left(q, V^2 \to 2V_{dc}V_n\right) = -\frac{\varepsilon S V_{dc}}{g^2}V_n \;. \qquad (27)$$

We will show in the rest of Supplementary Note 2 that, if we replace in the simplified equation of motion (Supplementary Equation 25) the linear noise force $F_n$ (Supplementary Equation 27), which possesses distinct oscillatory properties (see the definition of $V_n$ in Supplementary Equations 16 to 20), by the white noise specified in the main text (and below), most important features of dynamics of the system (Supplementary Equation 25) paradoxically remain almost unaffected, provided the system does not go in the phase space too far beyond a vicinity of the stable state of the resonator in the absence of noise. In particular, this relates to the dynamics determining the spectrum of fluctuations.

In intuitive terms, this paradox is explained as follows. The oscillator is almost linear and therefore the involved range of its eigenfrequencies is very narrow for relevant values of noise intensity, which is why its reaction to external forces has a strongly frequency-selective nature i.e. sharply weakens as a frequency of an external force goes beyond the aforementioned relevant narrow range of eigenfrequencies. Despite the spectrum of the original noise (Supplementary Equation 27) is narrow in comparison with its central frequency, it nevertheless fully covers the yet more narrow relevant range of eigenfrequencies and, moreover, includes also relatively broad ranges below and above the latter. That is why an addition to its spectrum of components of a comparable magnitude which lie even farther from the relevant range of eigenfrequencies cannot significantly affect the system. This in particular concerns the strictly white noise of such intensity at which the spectrum of the white noise approximately coincides with the spectrum of the original narrow-band noise (Supplementary Equation 27) in the relevant range of eigenfrequencies of the system i.e. in the close vicinity to $\omega_1$, which approximately corresponds to the maximum of the spectrum.

Let us present a rigorous substantiation of this important property. To this end, we transform in the equation of motion (Supplementary Equation 25) to energy and slow angle[3] (one could use other slow variables, e.g. those exploited in [4], too but we choose these since the angle is immediately related via $\omega(E)$ to the spectrum of involved eigenoscillations). More concretely, we first present a single 2$^{nd}$-order differential equation of motion as the equivalent system of two 1$^{st}$-order differential equations for $q$ and $p \equiv \dot{q}$, transform to energy $E$ and angle $\varphi$[2, 3], and take into account that the nonlinearity is weak which is why the dependence of $q$ and $p$ on $E$ and $\varphi$ may be approximated by that for the harmonic oscillator, i.e.

$$q - q_{eq}^{(n)} \approx \frac{\sqrt{2E/m}}{\omega_1} \cos(\varphi), \qquad p \approx \sqrt{2mE} \sin(\varphi) \tag{28}$$

($q_{eq}^{(n)}$ is the equilibrium (local minimum) position for $U^{(n)}(q)$ (Supplementary Equation 26): it slightly differs from $q_{eq}$).

Then we transform from the angle to the slow angle,

$$\tilde{\varphi} = \varphi - \omega_1 t \;, \tag{29}$$

and neglect in the resulting equation fast-oscillating terms, in accordance with the averaging method[1]. Ultimately, we obtain the following stochastic equations for energy and slow angle:

$$\dot{E} = -2\Gamma E - N_E(t),$$
$$\dot{\tilde{\varphi}} = \omega(E) - \omega_1 - N_{\tilde{\varphi}}(t), \tag{30}$$

where $N_E$ and $N_{\tilde{\varphi}}$ denote the following parametric noise terms:

$$N_E(t) = \frac{\varepsilon S V_{dc}}{g^2} \left(\frac{2E}{m}\right)^{\frac{1}{2}} \frac{\sin(\tilde{\varphi}) N_c(t) + \cos(\tilde{\varphi}) N_s(t)}{2},$$
$$N_{\tilde{\varphi}}(t) = \frac{\varepsilon S V_{dc}}{g^2} (2mE)^{-\frac{1}{2}} \frac{\cos(\tilde{\varphi}) N_c(t) - \sin(\tilde{\varphi}) N_s(t)}{2}. \tag{31}$$

The deterministic part of the dynamics of the system (Supplementary Equations 30 to 31) is slow in the sense that characteristic time-scales at which $E$ and $\tilde{\varphi}$ significantly change greatly exceed the correlation time of noise $t_{cor}$ (Supplementary Equation 23). Therefore, $d(t-t')$ in (Supplementary Equation 19) may be approximated by the delta-function in the context of main features of the dynamics (Supplementary Equations 30 to 31) (in particular, this concerns the spectrum of fluctuations):

$$d(t-t') \approx \delta(t-t'). \tag{32}$$

Since pair correlators of noise terms in the dynamic equations (Supplementary Equations 30 to 31) are $\delta$-correlated in the approximation (Supplementary Equation 32), the non-stationary probability density $W(E,\tilde{\varphi},t)$ obeys the Fokker-Planck equation (FPE)[5],

$$\frac{\partial W}{\partial t} = \hat{L}_{\text{FP}} W, \tag{33}$$

where the Fokker Planck operator $\hat{L}_{\text{FP}}$ can be presented in the form

$$\hat{L}_{\text{FP}} \equiv -\frac{\partial}{\partial E} D_E - \frac{\partial}{\partial \tilde{\varphi}} D_{\tilde{\varphi}} + \frac{\partial^2}{\partial E^2} D_{EE} + \frac{\partial^2}{\partial E \tilde{\varphi}}(D_{E\tilde{\varphi}} + D_{\tilde{\varphi}E}) + \frac{\partial^2}{\partial \tilde{\varphi}^2} D_{\tilde{\varphi}\tilde{\varphi}}, \tag{34}$$

where the drift and diffusion coefficients are fully determined by the form of the deterministic terms in the equation of motion (Supplementary Equations 30 to 31) and by the multipliers of $\delta$-functions in the pair correlators of the corresponding stochastic terms. Applying the general formulas[5] to the stochastic system (Supplementary Equations 30 to 31) with noises $N_c(t)$ and $N_s(t)$ given by Supplementary Equations 17 to 19 where $d(t-t')$ is approximated by the $\delta$-function, we obtain the following expressions for the drift and diffusion coefficients:

$$\begin{aligned}
D_E &= -2\Gamma E + \frac{D}{2m}, & D_{\tilde{\varphi}} &= \omega(E) - \omega_1, \\
D_{EE} &= \frac{DE}{2m}, & D_{E\tilde{\varphi}} &= D_{\tilde{\varphi}E} = 0, & D_{\tilde{\varphi}\tilde{\varphi}} &= \frac{D}{8mE},
\end{aligned} \tag{35}$$

where $D$ is an intensity of the linear noise force $F_n$:

$$\langle F_{\rm n}(t)F_{\rm n}(t')\rangle = 2Dd(t-t')\cos[\omega_{\rm l}(t-t')], \qquad D \equiv \left(\frac{\varepsilon SV_{\rm dc}}{g^2}\right)^2 I_{\rm n}. \qquad (36)$$

Let us introduce the strictly white (i.e. strictly $\delta$-correlated) noise $V_{\rm n}^{(w)}$ the intensity of which is twice smaller than the intensity of the original noise $V_{\rm n}$ given by Supplementary Equation 16:

$$\langle V_{\rm n}^{(w)}(t)V_{\rm n}^{(w)}(t')\rangle = 2I_{\rm n}^{(w)}\delta(t-t'), \qquad I_{\rm n}^{(w)} = \frac{I_{\rm n}}{2}. \qquad (37)$$

Let us now replace the original noise $V_{\rm n}$ in Supplementary Equation 27 for $F_{\rm n}$ by $V_{\rm n}^{(w)}$. In terms of the linear noise force, it means that the real force $F_{\rm n}$ in the equation of motion given by Supplementary Equation 25 is replaced by the white noise $F_{\rm n}^{(w)}$ of the twice smaller intensity:

$$\langle F_{\rm n}^{(w)}(t)F_{\rm n}^{(w)}(t')\rangle = 2D^{(w)}\delta(t-t'), \qquad D_{\rm w} = \frac{D}{2} \equiv \left(\frac{\varepsilon SV_{\rm dc}}{g^2}\right)^2 \frac{I_{\rm n}}{2}. \qquad (38)$$

Let us first, similarly to the case of the real noise (Supplementary Equation 16), transform in the equation of motion (Supplementary Equation 25) to energy and slow angle. The averaging over fast oscillations in the system is a more subtle issue as compared to the case of the noise (Supplementary Equation 16). In the latter case, the correlation time of noise greatly exceeds the period of fast oscillations and therefore the averaging may be applied immediately to the stochastic terms within the equation of motion. In contrast, the correlation time of the strictly white noise is zero and therefore the period of "fast" oscillations is slow as compared to it. Thus, the averaging of stochastic terms would be invalid. In order to correctly eliminate fast oscillations within the stochastic motion, we need first to transform to the description within the FPE while the averaging over fast oscillations should be done afterwards: the FPE is a deterministic equation and difficulties related to stochastic equations do not arise here. After the averaging is carried out in such a way, the resulting averaged FPE exactly coincides with the FPE relevant for stochastic equations being a result of the averaging of the equations of motion (Supplementary Equations 30 to 31), i.e. the drift and diffusion coefficients coincide after the averaging with those given in Supplementary Equation 35. It follows from this that main statistical properties of the two systems almost coincide: a relative difference is of the order of a ratio of period of eigenoscillation to the correlation time i.e. $2\pi/(\omega_1 t_{cor}) \ll 1$ (in particular, this concerns the spectrum of fluctuations). At the same time, it is worth noting that the equation of motion (Supplementary Equation 25) itself is a result of a neglect by small terms proportional to $D$ and an accurate estimate of some of them requires a more sophisticated white-noise approximation than that of the main term, but we do not take into account these small terms in the present work.

We emphasize that the spectrum of the idealized (white) voltage noise $V_n^{(w)}$ given in Supplementary Equation 37 coincides with the maximum value of that of the real voltage noise (Supplementary Equations 16 to 20) (Supplementary Figure 1). It is the most general requirement for a white noise to adequately mimic an original narrow-band noise: this requirement does not depend either on a concrete general form of the original noise or on a concrete shape of a correlation function $d(t)$ of its slow component.

Supplementary Note 3. Theoretical calculation of the spectrum of fluctuations.

The spectrum of fluctuations of a given dynamical variable is commonly defined as the half-Fourier transform of a correlation function of this variable [3-7],

$$\tilde{Q}(\omega) \equiv \frac{1}{\pi} \text{Re}\left[\int_0^\infty dt Q(t)\exp(-i\omega t)\right]. \tag{39}$$

and, often, a relevant dynamical variable is a generalized coordinate[3, 4, 7, 8] $q$. So, let us explicate the definition of the coordinate correlation function $Q(t)$. To the best of our knowledge, the most general definition existing in the literature is the following[3, 8]:

$$Q(t) = \lim_{\tau_1 \to \infty} \frac{1}{2\tau_1} \int_{-\tau_1}^{\tau_1} d\tau \left\langle \left(q(\tau+t) - \langle q(\tau+t)\rangle\right)\left(q(\tau) - \langle q(\tau)\rangle\right)\right\rangle, \tag{40}$$

where $\langle q(\xi)\rangle$ is a value of $q$ averaged over a statistical distribution of the system states at a given instant $\xi$ while the outer brackets $\langle \ldots \rangle$ mean an averaging over a statistical distribution of the system states both at the "initial" instant $\tau$ and at the "final" instant $\tau + t$. Consider a conditional probability density $\tilde{W}(q, p, \tau+t | q_0, p_0, \tau)$: the notation means that, if the system has coordinate $q_0$ and momentum $p_0$ at a given instant $\tau$, then the probability for the coordinate and momentum of the system at a given later instant $\tau+t$ to lie within the infinitesimally narrow intervals $[q, q+dq]$ and $[p, p+dp]$ respectively is equal to $\tilde{W}(q, p, \tau+t | q_0, p_0, \tau) dq dp$. There are classes of systems where, as time $t$ increases, $\tilde{W}$ approaches a stationary distribution independent of the initial conditions: $\tilde{W}_{st}(q,p) \equiv \lim_{t \to \infty}\left[\tilde{W}(q, p, \tau+t | q_0, p_0, \tau)\right]$. In such a case, the expression (Supplementary Equation 40) for the correlation function can be presented in a more explicit form:

$$Q(t) = \int_{-\infty}^{\infty}\int_{-\infty}^{\infty}\int_{-\infty}^{\infty}\int_{-\infty}^{\infty} dq dp dq_0 dp_0 \tilde{W}_{st}(q_0, p_0) \tilde{W}(q, p, t | q_0, p_0, 0)\left[(q-\langle q\rangle)(q_0 - \langle q\rangle)\right], \tag{41}$$

where $\langle q \rangle$ is the value of $q$ averaged over the stationary distribution,

$$\langle q \rangle \equiv \int_{-\infty}^{\infty}\int_{-\infty}^{\infty} dq dp \tilde{W}_{st}(q,p) q. \tag{42}$$

The broadest and most important class of systems possessing a stationary distribution is formed by systems moving in a potential field which goes to infinity as the coordinate goes to plus or minus infinity while being subject to a linear friction and an additive white noise with a constant intensity[5]:

$$\dot{p} = -2\Gamma p - \frac{dU(q)}{dq} + f_n(t),$$

$$\dot{q} = p, \tag{43}$$

where

$$\langle f_n(t) \rangle = 0, \qquad \langle f_n(t) f_n(t') \rangle = 4mk_B T\Gamma \delta(t-t'), \tag{44}$$

where $k_B$ is the Boltzmann constant while $T$ has a meaning of an effective temperature.

Such a system possesses the Gibbsian stationary distribution:

$$\tilde{W}_{st}(q,p) = \tilde{W}_{Gibbs}(E) \equiv \frac{1}{Z}\exp\left(-\frac{E}{k_B T}\right),$$

$$E \equiv E(q,p) = U(q) + \frac{p^2}{2m}, \tag{45}$$

$$Z = \left\{\iint dq dp \exp\left(-\frac{E}{k_B T}\right)\right\}^{-1}.$$

Before proceeding to our case, we need to generalize the definition of the fluctuation spectrum. Let a system of interest be metastable [5, 9] rather than truly stable. For the sake of simplicity, let us restrict the further discussion to potential systems i.e. those of the type (Supplementary Equation 43) but a potential field $U(q)$ may be arbitrary and noise $f_n(t)$ may not necessarily be white. A potential of a system possessing a metastable state necessarily includes at least one barrier so that, if the system exceeds the barrier, then it escapes from the metastable state for a long time or even forever. If the system is subject to noise, then the energy required for the escape is gained by the system from noise. If noise intensity is sufficiently small, then the mean escape time $t_{esc}$ [3, 5, 8-14], i.e. the average time during which the optimal (large) fluctuation [3, 10-13] required for the escape occurs, is exponentially (activation-like) large as compared both to a time-scale $t_{qs}$ during which a quasi-stationary distribution is established in the vicinity of the metastable state and to a time-scale $t_{cd}$ at which a correlation of motion in the vicinity of the metastable state substantially decays. Therefore, if to replace the upper integration limit in the definition of the spectrum (Supplementary Equation 39) by a large time $t_1$ from the range limited by the strong inequalities $\max\{t_{qs}, t_{cd}\} \ll t_1$ and $t_1 \ll t_{esc}$ while doing a statistical averaging in $Q(t)$ (Supplementary Equation 40) over a quasi-stationary distribution, then the integral

$$\tilde{Q}(\omega) \equiv \frac{1}{\pi} \mathrm{Re}\left[\int_0^{t_1} dt Q(t) \exp(-i\omega t)\right] \qquad (\max\{t_{qs}, t_{cd}\} \ll t_1 \ll t_{esc}) \qquad (46)$$

characterizes fluctuations in the vicinity of the metastable state analogously to how $\tilde{Q}(\omega)$ (Supplementary Equation 39) characterizes fluctuations in the vicinity of a stable state.

The spectrum measured by us in this work should be understood just in the way described in the previous paragraph since our resonator in the relevant range of $\tilde{V}_{dc}^2$ is a metastable system for which $t_{esc}$ is very large if the noise intensity lies in the relevant range indicated in Eq. (6). Let us demonstrate this explicitly. The phase plane of the noise-free system described by Eqs. (1)-(3) with $V^2 = \tilde{V}_{dc}^2$ (while $\tilde{V}_{dc}^2$ lies within the range indicated in Eq. (6)) possesses both a stable and unstable stationary states: respectively ($q = q_{eq} \approx \tilde{v}g, \dot{q} = 0$), which corresponds to the bottom of the corresponding potential well of $U^{(n)}(q, \lambda = 0) \equiv U(q) = U_s(q) + U_e(q)$ (Supplementary Equation 26), and ($q = q_b, \dot{q} = 0$), which corresponds to the top of the barrier of $U(q)$ situated at

$$q_b \approx g\left(1 - \sqrt{\tilde{v}}\right). \qquad (47)$$

The height of the barrier is:

$$\Delta U \equiv U(q_b) - U(q_{eq}) \approx \frac{m\omega_s^2 g^2}{2}. \qquad (48)$$

As $q$ increases beyond $q_b$, the potential $U(q)$ sharply decreases and ultimately drops to $-\infty$ as $q$ approaches $g$: if the top plate gets into such a close vicinity of the electrode, then it most likely sticks to it soon after that.

As soon as noise (e.g. random force $f_n(t)$ in Supplementary Equation 43) is added, the originally stable state in the bottom of the potential well becomes metastable since there is a non-zero probability $P$ for the large fluctuation which transfers the resonator over the barrier to occur. If noise is weak, then the dependence of this probability on the noise intensity $D_w$ is activation-like (cf. [3, 5, 8-14]):

$$P \propto \exp\left(-\frac{S}{D_w}\right), \qquad (49)$$

where $S$ is called action. Consider for example the system (Supplementary Equation 43) where the noise force $f_n(t)$ may generally speaking be arbitrary. If the correlation time of noise is much smaller than the period of natural oscillations $2\pi/\omega_1$ and the noise is additive (independent of $q$ and $p$, apart from being linear) i.e. if $f_n(t)$ is well approximated by Supplementary Equation 44, then action is equal to $2m\Gamma\Delta U$ so that the absolute value of the exponent in Supplementary Equation 49 is equal to $\Delta U/k_B T$ (cf. [3, 9, 10, 12]). However neither of the two above conditions is satisfied. On the one hand, the noise in our original equation of motion given by Eqs. (1)-(4) can be considered as independent of $q$ only in the vicinity of the metastable state, where the multiplier $(1-q/g)^{-1}$ can be approximated by 1, while the latter approximation is invalid in the major part of the energy range relevant to the escape i.e. where $E-U(q_{eq}) \sim \Delta U$: owing to this $q$-dependence, the effective noise intensity grows as energy increases (and even diverges as $E-U(q_{eq})$ approaches $\Delta U$), that leads to the decrease of action and therefore to the exponentially large increase of $P$ (Supplementary Equation 49). If noise correlation time is much smaller than the period of natural oscillation $2\pi/\omega_1$, the exponent in $P$ (Supplementary Equation 49) could be shown to be smaller of the conventional value $\Delta U/k_B T$ by the factor $7/2$. But the correlation time is, on the contrary, much larger than $2\pi/\omega_1$. Moreover, given that the eigenfrequency $\omega(E)$ goes to zero as energy $E$ approaches $U(q_b)$, our noise (the spectrum of which concentrates near the frequency $\omega_1$) is strongly non-resonant with eigenoscillations in the energy range close to the barrier value $U(q_b)$. Previous theoretical analysis (see e.g. [3, 10, 12]) showed that the optimal fluctuation which yields the conventional exponent $-\Delta U/k_B T$ in $P$ (Supplementary Equation 49) for the system (Supplementary Equations 43 to 44) is equal to $f_n^{(opt)}(t) = 2\Gamma p$: it results in the most probable escape path being time-reversed to the noise-free relaxational path from the top of the potential barrier into the bottom of the potential well [3, 10, 12]. Given that $\Gamma/\omega_1 \ll 1$, this relaxational path represents almost ideal eigenoscillations with the frequency $\omega(E(t))$ while $E(t)$ slowly relaxes from $U(q_b)$ to $U(q_{eq})$. Therefore the aforementioned optimal fluctuation oscillates in the energy range close to $U(q_b)$ with the slowly decreasing frequency $\omega(E(t)) \ll \omega_1$ while $E(t)$ slowly (with a characteristic time-scale $\omega_1/(\Gamma\omega(E(t)))$) increases within this range as time goes. Unlike the white noise (Supplementary Equation 44), the spectrum of the noise in our experiment does not contain so small frequencies and therefore such realization merely does not exist for it. Obviously, those realizations which do exist yield much larger action. This effect is much stronger than the aforementioned tendency to an increase of action owing to the $q$-dependence of the noise force. That is why, altogether $S/D_w$ greatly exceeds $\Delta U/k_B T$ and, as a result, the escape probability is vanishingly small provided the ratio $\Delta U/k_B T$ is large (or even moderate). Such conclusion conforms to our experiments: we did not observe any sticking of the plates when the noise intensity was within the range at which the noise-induced spectral narrowing took place.

Thus, either experimental or theoretical calculation of the fluctuation spectrum for our resonator in the relevant range of noise intensity does not differ in practice from cases of strictly stable systems. For example, while doing theoretical calculation, we may formally replace the unstable part of the potential function $U(q)$ (i.e. that for $q > q_b$) by any growing function: results for the spectrum near the natural frequency are not sensitive to such a replacement to an exponential accuracy.

As it is shown in Supplementary Note 2, the stochastic dynamics of the resonator in the presence of a resonant narrow-band noise component of voltage, e.g. of the form in Supplementary Equations 16 to 20, is well described within the additive linear noise approximation (Supplementary Equation 25) and, moreover, if the real noise $F_n$ is replaced by the white noise $F_n^{(w)}$ of a twice smaller intensity as explicitly defined in Supplementary Equation 38, main statistical features of the dynamics in the vicinity of the stable state (just such features determine the fluctuation spectrum in the range of the natural frequency of the resonator) remain almost the same. Allowing for this and for the above discussion of the equivalence to an exponential accuracy between the quasi-stationary distribution in the metastable system and the stationary distribution in the auxiliary stable system, we conclude that the resonator driven by the noise (Supplementary Equations 16 to 20) possesses a quasi-steady distribution close to the Gibbsian stationary distribution [3, 5-7]:

$$\tilde{W}_{st}(q,p) = \tilde{W}_{Gibbs}(E) \equiv \frac{1}{Z}\exp\left(-\frac{E}{D_w/(2m\Gamma)}\right),$$

$$E \equiv E(q,p) = U^{(n)}(q) + \frac{p^2}{2m} \approx \frac{m\omega_1^2 q^2}{2} + \frac{p^2}{2m}, \qquad (50)$$

$$Z = \left\{\iint dq dp \exp\left(-\frac{E}{D_w/(2m\Gamma)}\right)\right\}^{-1} \approx \frac{4\pi\Gamma}{\omega_1 D_w},$$

where $D_w$ is defined in Supplementary Equation 38 in Supplementary Note 2.

One more remarkable property of systems with the white noise is that the time evolution of the non-stationary probability density $\tilde{W}$ obeys the Fokker-Planck equation [3-5, 7] (FPE) i.e certain partial differential equation of the second order (cf. Supplementary Equations 33 to 35 in Supplementary Note 2). Still, its solution is complicated, even in case of a weak nonlinearity and even numerically. The latter is especially true in our weakly nonlinear case, where pronounced characteristic features of the fluctuation spectrum may result from rather subtle differences of an evolution of the probability density from that in the purely linear case and from that in a conventional weakly nonlinear case. Fortunately, a powerful method for the FPE solution and the calculation of fluctuation spectra on its base was developed for underdamped cases earlier: first for monostable potentials[7] and then for multistable ones[15]. As a result, a solution of the complicated partial differential equation and a heavy integration over a double phase space and time is reduced to a solution of a relatively simple ordinary differential equation with boundary conditions and a simple integration just over energy. One can find details in the aforementioned papers[7, 15] or in the review[3]. Here, we just refer to the relevant general result, presenting it in notations used in (or just relevant to) the present paper. The spectrum of fluctuations reads as follows:

$$\tilde{Q}(\Omega) \approx 2\,\text{Re}\left[\int_0^{E_{up}} dE \frac{1}{\omega^{(n)}(E)} \left(q_1^{(n)}(E)\right)^* W_1(E,\Omega)\right], \qquad (51)$$
$$|\Omega - \omega_1| \ll \omega_1,$$

where the superscript $\ldots^{(n)}$ here and thereafter means a modification of a given quantity owing to the slight modification of the original potential by the quadratic noise (see $U^{(n)}(q)$ Supplementary Equation 26), $E_{up}$ is a rather arbitrarily chosen energy which greatly exceeds an average energy in the quasistationary state for a given noise intensity $D_w$ but being lower than the potential barrier height $\Delta U^{(n)}$:

$$\langle E \rangle \ll E_{up} < \Delta U^{(n)},$$
$$\langle E \rangle \approx \frac{D_w}{2m\Gamma}, \qquad (52)$$

$\left(q_1^{(n)}(E)\right)^*$ is a complexly conjugated quantity to the first-order Fourier component of coordinate as function of energy and phase for the conservative system with the noise-modified potential $U^{(n)}(q)$,

$$q_1^{(n)}(E) = \frac{1}{2\pi} \int_0^{2\pi} d\varphi \exp(-in\varphi) q^{(n)}(E, \varphi),$$

$$\left(q_1^{(n)}(E)\right)^* = \frac{1}{2\pi} \int_0^{2\pi} d\varphi \exp(in\varphi) q^{(n)}(E, \varphi),$$

(53)

and $W_1(E, \Omega)$ is a solution of a boundary problem which reads in a compact form as follows:

$$-i\left[\Omega - \omega^{(n)}\right] W_1 = 2\Gamma \left\{ \left[ 1 + \frac{\overline{(p^{(n)})^2}}{m} \frac{d}{dE} \right] \left[ 1 + \frac{D_w}{2m\Gamma} \frac{d}{dE} \right] - \frac{D_w}{2m\Gamma} \left(\omega^{(n)}\right)^2 \overline{\left(q_E^{(n)}\right)^2} \right\} W_1 + q_1^{(n)} \tilde{W}_{\text{Gibbs}},$$

$$W_1(E = 0, \Omega) = 0, \qquad W_1(E = E_{\text{up}}, \Omega) = 0,$$

$$\frac{D_w}{2m\Gamma} \ll \Delta U,$$

(54)

where $\omega^{(n)} \equiv \omega^{(n)}(E)$ is the eigenfrequency vs. energy for the noise-modified dc voltage (see Supplementary Equation 25 in Supplementary Note 2),

$$\omega^{(n)}(E) = \omega(E) - \frac{D_w f_{\text{bd}} g}{\sqrt{\ln(2)/\pi} m\omega_s \varepsilon S V_{\text{dc}}^2} \left( 1 + \frac{3}{mg^2 \omega_s^2} E \right)$$

(55)

(in the noise-induced part of $\omega^{(n)}(E)$, terms $\propto E^k$ with $k \geq 2$ are negligible in the relevant range of $E$, which is why they are omitted in Supplementary Equation 55) with the bandwidth $f_{\text{bd}}$ equal to 741 Hz for the conventional case (Supplementary Figure 1) and to 298 Hz for the zero-dispersion case, the over-bar $\overline{\ldots}$ in Supplementary Equation 54 means the averaging over the phase i.e.

$$\overline{\left(p^{(n)}\right)^2} \equiv \overline{\left(p^{(n)}\right)^2}(E) = \frac{1}{2\pi} \int_0^{2\pi} d\varphi \left(p^{(n)}(E,\varphi)\right)^2, \tag{56}$$

$$\overline{\left(q_E^{(n)}\right)^2} \equiv \overline{\left(q_E^{(n)}\right)^2}(E) = \frac{1}{2\pi} \int_0^{2\pi} d\varphi \left(\frac{\partial q^{(n)}(E,\varphi)}{\partial E}\right)^2, \tag{57}$$

and $\tilde{W}_{\text{Gibbs}} \equiv \tilde{W}_{\text{Gibbs}}(E)$ is the quasi-stationary distribution given in Supplementary Equation 50.

For our system, the ordinary differential equation in Supplementary Equation 54 can be further simplified. In the relevant range of energy $E$ and noise intensity $D$, the conservative approximation is very close to the harmonic oscillator in the sense that the relative deviation of the coefficients $\left(\omega^{(n)}(E)\right)^2$, $\overline{\left(p^{(n)}\right)^2}(E)$, $\overline{\left(q_E^{(n)}\right)^2}(E)$, $q_1^{(n)}(E)$ and $\tilde{W}_{\text{Gibbs}}(E)$ from those in the harmonic approximation is negligible (being $\lesssim 10^{-5}$) and therefore all these coefficients in the right-hand side of the differential equation in Supplementary Equation 54 may be replaced by their harmonic approximations. On the contrary, the deviation of $\omega^{(n)}(E)$ from $\omega_1$ in the imaginary left-hand side of the equation plays the crucial role: it is just the behavior of this deviation as function of $E$ that determines main features of the fluctuational spectrum. Using the harmonic approximations for the coefficients in the right-hand side of the differential equation in Supplementary Equation 54 and introducing proper normalizations of relevant quantities and parameters, we can present the procedure of finding the fluctuation spectrum in the following rather simple way. The spectrum is equal to

$$\tilde{Q}(\Omega) = \frac{\langle E \rangle}{2m\Gamma\omega_1^2} \int_0^{E_{\text{up}}/\langle E \rangle} dx \sqrt{x} \, \text{Re}[y(x)], \tag{58}$$

where $x$ is a dimensionless variable having a meaning of energy $E$ normalized by the average energy $\langle E \rangle$ (given in Supplementary Equation 52), and $y(x)$ is a dimensionless complex function being a solution of the following ordinary differential equation

$$x\frac{d^2 y}{dx^2} + (1+x)\frac{dy}{dx} + \left(1 - \frac{1}{4x} + i\frac{\Omega - \omega^{(n)}(E = x\langle E \rangle)}{2\Gamma}\right)y = -\frac{\sqrt{x}}{2\pi}\exp(-x), \quad (59)$$

which satisfies the boundary conditions

$$y(x=0) = y(x = E_{up}/\langle E \rangle) = 0. \quad (60)$$

Supplementary Note 4. Allan deviation of the oscillator in the zero-dispersion regime

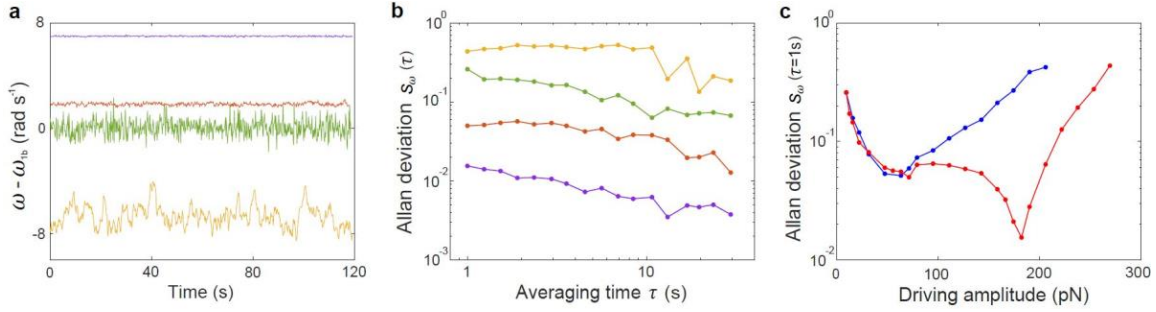

**Supplementary Figure 2** Frequency fluctuations and Allan deviation in the zero-dispersion regime. **a** Measured frequency of the self-sustained oscillations driven with feedback for $V_{dc}$ = -1.59V at driving amplitudes of 22.2 pN (green), 71.3 pN (brown), 182 pN (purple) and 269 pN (yellow). **b** Allan deviation of the data in **a** as a function of the averaging time. **c** Allan deviation of the frequency of self-sustained oscillations at the given $\tau = 1s$ versus driving amplitude for oscillators in the conventional regime (blue) and the zero-dispersion regime (red). The Allan deviations are calculated using the same recorded frequencies as those in Fig. 4d of the main text.

In Fig. 4d of the main text, the resonator is driven with feedback using a phase locked loop. The frequency for each data point is recorded over a duration of 120 s. Supplementary Figure 2a shows the recorded frequency as a function of time for four different driving amplitudes of 22.2 pN, 71.3 pN, 182.1 pN and 269 pN. In addition to characterizing the frequency stability using the standard deviation of frequency, we perform the analysis using the Allan deviation:

$$s_\omega(\tau) = \sqrt{\sum_{i=1}^{N-1}(\overline{\omega}_{i+1}^\tau - \overline{\omega}_i^\tau)^2 \Big/ 2(N-1)} \qquad (61)$$

where N is the maximum number of non-overlapping time intervals of duration $\tau$ in the total recording time of 120 s. $\bar{\omega}_i^\tau$ represents the recorded frequency averaged over the $i^{th}$ interval. Supplementary Figure 2b shows that $s_\omega(\tau)$ decreases with $\tau$, supporting the notion that the effect of long term frequency drift is small for the recording time. Supplementary Figure 2c plots the Allan deviation $s_\omega$ ($\tau$ = 1s) as a function of driving amplitude for the zero dispersion and conventional nonlinear regimes, using the same records of frequency as in Fig. 4d in the main text. The improvement of the Allan deviation in the zero dispersion regime over the convention regime is about a factor of 3.6.

## Supplementary References